\documentclass[a4,twocolumn]{revtex4}
\usepackage{graphicx}
\usepackage{epstopdf}
\usepackage{longtable}
\usepackage{dcolumn}
\usepackage{bm}
\usepackage{amsmath}
\usepackage{amssymb}
\usepackage{color}
\usepackage[colorlinks=true, linkcolor=blue, menucolor=blue, citecolor=blue, dvipdfm]{hyperref}

\begin{document}

\title{Unconventional approach to orbital--free density functional
theory \\ derived from a model of extended electrons.}

\author{Werner A. Hofer}
\affiliation{Department of Physics, University of Liverpool\\
L69 3BX Liverpool, United Kingdom}

\begin{abstract}
An equation proposed by Levy, Perdew and Sahni in 1984 [PRA 30, 2745
(1984)] is an orbital--free formulation of density functional
theory. However, this equation describes a bosonic system.
Here, we analyze on a very fundamental level, how this equation
could be extended to yield a formulation for a general fermionic
distribution of charge and spin. This analysis starts at the
level of single electrons and with the question, how spin actually
comes into a charge distribution in a non-relativistic model. To
this end we present a space-time model of extended electrons, which
is formulated in terms of geometric algebra. Wave properties of the
electron are referred to mass density oscillations. We provide a
comprehensive and non-statistical interpretation of wavefunctions,
referring them to mass density components and internal field
components. It is shown that these wavefunctions comply with the
Schr\"odinger equation, for the free electron as well as for the
electron in electrostatic and vector potentials. Spin-properties of
the electron are referred to intrinsic field components and it is
established that a measurement of spin in an external field yields
exactly two possible results. However, it is also established that
the spin of free electrons is isotropic, and that spin-dynamics
of single electrons can be described by a modified Landau-Lifshitz equation.
The model agrees with the results of standard theory concerning the hydrogen
atom. Finally, we analyze many-electron systems and derive a set of
coupled equations suitable to characterize the system without any
reference to single electron states. The model is expected to have
the greatest impact in condensed matter theory, where it allows to
describe an $N$-electron system by a many-electron wavefunction
$\Psi$ of four, instead of $3N$ variables. The many-body aspect of a
system is in this case encoded in a bivector potential.
\end{abstract}

\maketitle

\section{Introduction}

Electronic structure simulations today underpin many models
constructed to account for experimental data. Most of these
simulations are based on an implementation of density functional
theory (DFT). DFT itself is theoretically justified by the
Hohenberg-Kohn theorems \cite{hk64}, even though a model based on
the electron density was suggested much earlier by Thomas and Fermi
\cite{thomas,fermi}. A combination of Thomas-Fermi model and
Hohenberg-Kohn DFT leads to the following description of a
many-electron system \cite{wang06},
\begin{equation}
\frac{\delta E[\rho]}{\delta \rho ({\bf r})} = \frac{\delta
T[\rho]}{\delta \rho({\bf r})} + v_{H}[\rho]({\bf r}) +
v_{ne}[\rho]({\bf r}) + v_{xc}[\rho]({\bf r}) = \mu.
\end{equation}
Here, $T[\rho]$ is the kinetic energy functional, depending on the
density of electron charge $\rho$, $v_H$ the Hartree potential or
the electron-electron repulsion, $v_{ne}$ the electron-nuclei
attraction, and $v_{xc}$ the exchange correlation potential. $\mu$
is the Lagrange multiplier due to the condition of charge
conservation, or the chemical potential.

While this equation is generally valid, a transferable and fully
general kinetic energy functional based on the density has so far
remained elusive. It is known that the functional for the
homogeneous electron gas is described exactly by the Thomas-Fermi
kinetic energy functional (TF), or \cite{thomas,fermi,wang99}
\begin{equation}
T_{TF}[\rho] = C_{TF} \langle \rho^{5/3}({\bf r})\rangle.
\end{equation}
For hydrogen atoms, it is described exactly by the von Weizs\"acker
functional (vW), or \cite{weiz35}
\begin{equation}
T_{vW}[\rho] = \frac{1}{8} \left\langle \frac{|\nabla \rho ({\bf
r})|^2}{\rho({\bf r})}\right\rangle.
\end{equation}
In present implementations of orbital-free DFT one either
interpolates between the TF and vW functionals for more general
density distributions \cite{wang99,perdew07}, or one bases the whole
calculation or parts of it on the Kohn-Sham (KS) method of DFT
\cite{wang06,ks65}. The reason, one tries to avoid the KS method as
much as possible is that the computational effort in this case
scales with the cube of $N$, the number of electrons, which limits
the number of atoms which can be included in the simulation to a few
thousand. However, most of the technologically interesting materials
e.g. in the semiconductor industry are very low doped; one dopant
per millions of atoms of the host matrix. In this case current
methods are unable to reliably simulate the material.

In a paper in 1984 Levy, Perdew and Sahni \cite{lps84} showed that
one should be able to describe the general density of a system by:
\begin{equation}\label{lps-01}
\left[-\frac{1}{2} \nabla^2 + v_{ext} + v_{eff} - \mu
\right]\rho^{1/2} = 0.
\end{equation}
Here, $v_{ext}$ is the external ionic potential, $v_{eff}$ is the
electron-electron repulsion combined with the exchange-correlation
potential, and $\mu$ is again the chemical potential, which can be
interpreted as a generalized and constant energy density. The
problem, which was left unsolved, is that the equation seems to
describe a bosonic system. Subsequently, Norman March and others
extended the equation to account for the fermionic nature of electrons.
They introduced a Pauli potential, which describes Pauli repulsion
between individual electrons in a system \cite{march86,march87,holas91}.

This was the starting point of our analysis of orbital free DFT some
years ago. The appeal of the equation is that it has the
same general form as the Schr\"odinger equation and that it only
includes the density, thus is an orbital--free formulation of DFT.
But it is necessary to extend the equation and to generalize it, so that it
can be used for arbitrary fermionic systems. This, as will be shown, can be
accomplished in a consistent manner. While previous approaches focussed
on the extension of the effective potential \cite{march86}, within the present framework we extend
the square root of the density to incorporate electron spin. The analysis necessary to
accomplish this task is very fundamental, since it has to start with
the question, how spin actually comes into a density distribution in
the non-relativistic limit. Such a fundamental analysis has to include not
only the structure of electrons, but also its interactions with
electromagnetic fields to establish, that the model is fully
consistent with the standard results. The structure of the paper is
as follows:

In Section \ref{single-electron} we review existing electron models
in view of their suitability for the tasks at hand and give four
postulates, which we think capture the requirements for a
comprehensive electron model. We also introduce the wavefunction
$\psi$ of single electrons as a multivector of three dimensional
geometric algebra.

In Section \ref{geometric-algebra} we give a short overview over
geometric algebra necessary for the presentation. As condensed
matter theorists will most likely not be acquainted with geometric
algebra, we think this section is necessary for an understanding of
the concepts.

In Section \ref{wave-properties} we present a motivation why
potential components of electrons in motion are necessary to attain
wave properties of moving electrons. These electromagnetic energy
components differ from the classical components previously
introduced by Lorentz and Abraham.

In Section \ref{wave-1} we present a consistent and local model of a
single-electron's density and spin components in geometric algebra
as well as a relation between these components and the wavefunction
of the electron.

In Section \ref{static} we determine the interactions of the
electron with external static fields and show how the density and
spin components change upon interaction.

In Section \ref{dynamic} we analyze the interaction of electrons and
photons and show that the effect on the electron's properties can be
mimicked by a calculation including only momenta and energies of
electrons and photons.

In Section \ref{hydrogen} we show that the model is consistent with
the standard model of hydrogen. However, we also present an
extension of the standard model in the non-relativistic regime and
including the spin of the electron.

In Section \ref{hydrogen-molecule} we present the model of hydrogen
molecules and extend Eq. (\ref{lps-01}) to an arbitrary density
distribution and including the spin state of electrons.

Finally, in Section \ref{discussion} we discuss the presented
concepts in view of fundamental issues in quantum mechanics and the
consistency of the model with existing experiments, in particular
the experiments on fundamental quantum mechanics. We also present an
outlook on future work to be undertaken to check on the method.

The guiding principle of this work is the 'radical conservative -
ism' of John Archibald Wheeler \cite{wheeler-1}: {\it Insist on
adhering to well-established physical laws (be conservative) but
follow those laws into their most extreme domains (be radical),
where unexpected insights into nature might be found}. We do not
attach any ontological claim to the model suggested. It is adopted
purely for its usefulness, as it greatly simplifies the description
of many-electron systems, in fact allowing for a coherent
theoretical model from the level of single electrons to, in
principle, macroscopic systems.

\section{Single electron models} \label{single-electron}

Generally, it has to be conceded that all attempts to ascribe a
reality to the electron, which goes beyond a point particle with an
intrinsic momentum and thus magnetic moment, have failed. The most
convincing arguments against an internal structure and for a
point-like electron come from single-electron diffraction and
high-energy scattering experiments. In the first case, the
point-like impacts of the single particles, gradually building up a
diffraction pattern, are one of the great paradigms in quantum
physics, which establish the existence of a guiding principle behind
the statistical events \cite{donati73}. Similarly, high--energy
scattering experiments are routinely used to check the predictions
of quantum field theory. So far, the agreement between experiment
and theory is better than one part in a billion \cite{gabrielse07}.

There have been attempts, most notably by Louis de Broglie, David
Bohm, and David Hestenes
\cite{bohm52,ldb71,val09,hest73,hest85,hest90}, to ascribe the
observed duality of the electron, its wave-features and its discrete
charge and mass without manifest volume, to a field-like construct.
The pilot-wave theory, originally due to Louis de Broglie, has even
gained renewed interest in recent years, see for example the article
by Antony Valentini in a recent Physics World issue \cite{val09}.
However, it is not clear, how these models can be reconciled with
the fundamental fields of interaction known in physics. These
interactions are all subject to relativity and thus do not allow for
action-at-a-distance. Ascribing a physical reality to the
wavefunction of an electron itself, while still retaining the
mechanical properties of a point particle, inevitably seems to lead
to a pilot-wave or Bohm-type theory, where the potentials themselves
become non-local \cite{bohm52}. It thus only shifts the central
problem, contained in the following question: ''What actually {\em
is} an electron, and which physical property describes its spin?''
to the question: ''What actually is a {\em quantum potential} and in
which way does it relate to fundamental interactions?'' In addition,
it remains unclear, in these models, how spin enters the description
of electrons, and how it interacts with magnetic fields. As this is
the fundamental problem to be solved in the present context, one
needs to go beyond the existing frameworks.

If one takes current experimental results at the atomic scale
seriously, then one arrives most likely at the conclusion of Don
Eigler \cite{eig02}: ''I don't believe in this wave-particle duality
... I think it's mostly just the left-over baggage of having started
off understanding the world in terms of particles and then being
forced, because of the quantum revolution, to think of the world in
terms of waves. Don't even think about them as particles. Electrons
are waves. And if you think of them in terms of waves, you will
always end up with the right answer. Always.'' A similar point of
view, based on a mathematical analysis of quantum mechanical
concepts, has recently been put forward by Hrovje Nikolic
\cite{nikolic07}, calling the wave--particle duality a "myth". In
addition, our most successful theory to describe solids, density
functional theory (DFT), even though it is based on second
quantization, still ascribes a reality only to the density of
electron charge and its field of interaction \cite{hk64}. If this is
the case, then a model of electrons could also be constructed, not
from mechanics and an elementary mass or charge, but from a
wave-like structure which interacts in the same way electrons do.

At this point one might observe that it is simply impossible to
account for the electron's spin within a classical model, i. e. a
model where spin is represented by a vector in real-space. However,
as Hestenes, and Doran and Lasenby have shown \cite{hest84,dor02},
the algebra of electron spin, described by the Pauli matrices, is
also obtained by a vector model within {\em Geometric Algebra}
(maybe better known as {\em Clifford Algebra}), where it results
from the general properties of rotations in three-dimensional space.
Along the same lines of thought one also arrives at the result that
the $\gamma$-matrices in the Dirac equation of the electron can be
seen as expressions of the geometric algebra of four-dimensional
space-time. From these facts it is possible to conclude that spin
can actually be seen as a ''classical'' geometric property of the
spacetime representation of the electron, which can be described by
the even elements of a multivector in geometric algebra
\cite{dor02}. If we choose to represent the wavefunction in a manner
which bears on these relations, then it can be written in the
following form:
\begin{equation}
\psi = \alpha + \sum_{i \ne j} {\bf e}_i \wedge {\bf e}_j \beta_{ij}
\qquad \alpha,\beta_{ij} \in R
\end{equation}
Here, the symbol $\wedge$ defines the ''wedge-product'' in geometric
algebra between the vectors ${\bf e}_{i(j)}$ of the
three-dimensional frame, which yields a ''directed-plane'' in space.
Incidentally, in a frame of three dimensions this wedge product can
be interpreted as the conventional cross-product $\times$ between
two vectors times a ''pseudoscalar'', which in turn can be seen as
the imaginary unit, or $i$. In three dimensional space the imaginary
component can be interpreted in three different ways : (i) It is
either a ''directed-plane'', which means it is due to the directions
of two independent vectors, or (ii) it is a vector, since every
directed-plane is dual to a vector, or (iii) it represents an angle
and an axis of rotation \cite{hest84,dor02}. We choose the first
interpretation in the following, because it allows for a
straightforward interpretation of the observed oscillations of
electron waves as charge density oscillations.

We shall start from some ideas developed more than ten years ago
\cite{hof98}, and show that they can be extended to yield a model of
electrons which is free of contradictions, in accordance with most
fundamental results obtained in quantum mechanics, shows that a
''wavefunction'' of the electron exists which must comply with the
Schr\"odinger equation, and which allows a representation of
electron properties within geometric algebra. The clear difference
to previous models, e.g. the ones by Louis de Broglie \cite{ldb71},
David Hestenes \cite{hest85} or Jaime Keller \cite{keller01} is that
(i) the relation between wavefunctions and physical variables like
mass density or field amplitudes is made explicit, and (ii) that it
is shown in a local and time dependent picture how external applied
fields directly translate into a change of the wave-properties of
the electron.

The clear disadvantage of the model is that it is unsuitable to
describe single-electron diffraction \cite{donati73}. In this
respect, the interpretation of the electron as a point particle and
the interpretation of the wavefunction as a probability measure
seems indispensable.

However, the model does have conceptual advantages in that it allows
for a direct and locally defined relationship between physical
properties and (single-electron) wavefunctions. This gain, we think,
makes it worthwhile to interpret electrons in this way. Moreover, it
allows for a very efficient formulation of many-electron problems.
There, it leads to the introduction of a conceptually new bivector
potential. This potential is necessary to account for
spin-properties in a correlated system. The description also remains
remarkably simple and is thus potentially very useful in the
development of numerical methods for the simulation of solid state
and molecular systems.

The model of electrons rests on four distinct postulates:
\begin{enumerate}
\item The wave properties of electrons are a {\em real} physical
property of electrons in motion.
\item Electrons in motion possess intrinsic electromagnetic potentials which
are vector-like.
\item The magnetic moment of electrons is a consequence of the orientation of these
electromagnetic potentials.
\item In equilibrium the energy density
throughout the space occupied by a single electron is a constant.
\end{enumerate}

It is a discerning feature of postulates that they cannot be derived
from phenomena. The same applies to these postulates. However, their
logical consequences have to be in line with experiments or a
generally applicable theoretical framework. Moreover, they should
make these consequences more transparent, i.e. they should lead to a
gain in understanding. Finally they should enhance our ability to
predict experimental results, i.e. they should lead to an extension
of theoretical methods. We think, all this applies to our
postulates, as we shall show presently.

\section{A short introduction to geometric
algebra}\label{geometric-algebra}

Most solid state physicists will not be familiar with geometric
algebra. This short introduction is modeled on an introductory paper
by Gull, Lasenby and Doran \cite{gull93}, which in our view provides
the easiest introduction to the topic. In a Gibbs vector algebra,
built on Euclidean geometry, we have two separate products between
vectors, an inner product ${\bf a} \cdot {\bf b}$, which is a
scalar:
\begin{equation}
{\bf a} \cdot {\bf b} = \alpha \in R,
\end{equation}
and an outer product ${\bf a} \times {\bf b}$, which is a vector:
\begin{equation}
{\bf a} \times {\bf b} = {\bf c} \in R^3.
\end{equation}
In three dimensions the vector product is equal to the wedge product
times the imaginary unit $i$, so that  ${\bf a} \times {\bf b} = - i
{\bf a} \wedge {\bf b}$.  In geometric algebra the two products are
combined in a Clifford product or the geometric product between
vectors.
\begin{equation}
{\bf a} {\bf b} = {\bf a} \cdot {\bf b} + {\bf a} \wedge {\bf b}
\end{equation}
The geometric product contains thus two parts: a scalar, which is
symmetric, i.e., it does not change its sign upon a change of the
order of ${\bf a}$ and ${\bf b}$, and a {\em bivector}, which is
antisymmetric:
\begin{eqnarray}
{\bf a} \cdot {\bf b} &=& \frac{1}{2}\left({\bf a}{\bf b} + {\bf
b}{\bf a}\right) = {\bf b} \cdot {\bf a}\nonumber \\
{\bf a} \wedge {\bf b} &=& \frac{1}{2}\left({\bf a}{\bf b} - {\bf
b}{\bf a}\right) = - {\bf b} \wedge {\bf a}
\end{eqnarray}
The product of two parallel vectors is a scalar, while the product
of two orthogonal vectors is a bivector, denoted by the wedge
product. The general product usually contains both, a scalar and a
bivector. The wedge product can be seen as a plane, more
specifically a plane, the orientation of which is given by the order
of the two vectors in the wedge product. In geometric algebra, this
is called a directed plane.

It is important to realize that the geometric product, which adds a
scalar and a bivector, will have properties of both, the scalar {\em
and} the bivector. This is exactly like the relation between the
real part and the imaginary part in a complex number. In two
dimensions it can be shown that the wedge product carries the
imaginary unit. The proof is simple. We assume two framevectors,
${\bf e}_1$ and ${\bf e}_2$, which are orthogonal. Then we get for
the square of the wedge product:
\begin{equation}
\left({\bf e}_1 \wedge {\bf e}_2\right)^2 = {\bf e}_1{\bf e}_2{\bf
e}_1{\bf e}_2 = - {\bf e}_1{\bf e}_1{\bf e}_2{\bf e}_2 = -1
\end{equation}
This means that the wedge product carries an additional unit, the
imaginary unit. A multivector, i.e., a sum of a scalar and a
bivector in two dimensions can therefore be mapped onto the complex
number space:
\begin{equation}
z = x + y{\bf e}_1{\bf e}_2 \equiv x + i y
\end{equation}
At a deeper level this property is related to the fact that a wedge
product is actually a rotator, or an element of geometric algebra
which rotates vectors. This can be seen by acting with a wedge
product on the frame vectors:
\begin{eqnarray}
({\bf e}_1{\bf e}_2) {\bf e}_1 &=& - {\bf e}_2{\bf e}_1{\bf e}_1 = -
{\bf e}_2 \nonumber \\
({\bf e}_1{\bf e}_2) {\bf e}_2 &=& {\bf e}_1{\bf e}_2{\bf e}_2 =
{\bf e}_1
\end{eqnarray}
Rotations are in most undergraduate physics courses described by
matrices. The possibility to describe rotations by geometric
products suggests a deeper link between geometric products and
matrices. In three dimensions a general multivector is described by
a scalar, a vector, a bivector and a trivector. The geometric
product between frame vectors has the following algebra:
\begin{eqnarray}
i &=& {\bf e}_1{\bf e}_2{\bf e}_3 \nonumber \\
{\bf e}_1{\bf e}_2 &=& i {\bf e}_3 \nonumber \\
{\bf e}_2{\bf e}_3 &=& i {\bf e}_1 \\
{\bf e}_3{\bf e}_1 &=& i {\bf e}_2 \nonumber \\
{\bf e}_i{\bf e}_i &=& 1 \nonumber
\end{eqnarray}
Here, $i$ is actually a pseudoscalar, or the highest geometric
product (the product of three independent vectors) in three
dimensional space. The algebra of three dimensional space is
equivalent to the algebra of the Pauli matrices, given by
\cite{gull93}:
\begin{eqnarray}
{\bf e}_i{\bf e}_j &=& \delta_{ij} + i \epsilon_{ijk} {\bf e}_k
\nonumber \\
 \hat{\sigma}_i \hat{\sigma}_j &=& I
\delta_{ij}+ i \epsilon_{ijk} \hat{\sigma}_k
\end{eqnarray}
The Pauli matrices $\hat{\sigma}_k$ in this case can be seen as
matrix representations of the 3-dimensional geometric algebra. A
similar relation between framevectors of a 4-dimensional geometric
algebra and Dirac's $\gamma$-matrices indicates that these matrices
also can be seen as an expression of 4-dimensional spacetime
\cite{dor02}.

From a general point of view it thus turns out that imaginary
components are quite frequently an indication of a geometric basis
behind our formulations in a given vector space. It will be seen
that this allows for a natural distinction between density related
and spin related components in the electron wavefunction.

\section{Wave properties and electrons}\label{wave-properties}

In this section we review the original motivation to postulate the
existence of field--like energy components for an electron in
motion. We show that if one assumes that wave properties of
electrons are similar to the wave properties of electromagnetic
fields, then the total energy of an electron will be different from
its kinetic energy. As stated in previous publications, this is the
only possibility to account for wave properties in a physical
picture \cite{hof98}. The approach in quantum mechanics, where one
assumes from the outset that the only energy component of an
electron must be its kinetic energy, leads invariably to the
conclusion that the wave properties of the electron cannot be real
which requires then to interpret them as related to the probability
density \cite{born26}. This has been pointed out some time ago.
However, here we start from the picture of waves developed in
electrodynamics, and shall extend this picture to encompass also
electrons in motion. In electrodynamics, the relation between the
wavelength $\lambda$ of an electromagnetic wave and its frequency in
vacuum $\nu$ is given by the dispersion relation:

\begin{equation}\label{lnu}
\lambda \nu = c
\end{equation}
Here, $c$ is the velocity of a photon in vacuum. For photons we also
must account for Planck's relation between the energy $E$ and the
frequency $\nu$ of a particle:

\begin{equation}\label{planck1}
E = h \nu
\end{equation}
Here, $h$ is the Planck constant, or $6.626 \times 10^{-34} Js$. For
electrons, we start with the de Broglie relation between wavelength
and momentum of an electron. This relation was first verified by
electron scattering experiments of Davisson and Germer in the 1920s
\cite{dav27} and is today checked routinely in many labs around the
world performing low energy electron diffraction experiments. It
states that:

\begin{equation}\label{broglie1}
\lambda = \frac{h}{m v_{el}}
\end{equation}
$m$ in this case is the mass of the electron, or $9.1 \times
10^{-31} kg$. To see, what happens, if we assume that the Planck and
de Broglie relations are both valid for an electron, we now take the
velocity $c$ in Eq. (\ref{lnu})  to be the velocity of the electron
$v_{el}$, and combine it with Eqs. (\ref{planck1}) and
(\ref{broglie1}). Then we find that the energy of the electron
should be twice its kinetic energy, since:

\begin{equation}\label{total_simple}
\lambda = \frac{v_{el}}{E/h} = \frac{h}{m v_{el}} \quad
\Longrightarrow \quad E = m v_{el}^2 = 2 \times E_{kin}
\end{equation}
If we assume that an electron has wave properties which are similar
to the wave properties of a photon, then we immediately arrive at
the result that its energy is not its kinetic energy alone. It is,
of course, not very plausible that its energy is double its kinetic
energy, since this would require in every energy balance of electron
acceleration or deceleration that a factor of two should
mysteriously show up in the balance sheet. This has never been
observed, so the conclusion is fairly safe that this cannot be
strictly correct. However, it illustrates a general point which will
be the basis of our electron model: wave properties related to
physical properties of single electrons (not a statistical manifold
of many electrons) are only possible if the electron possesses more
than just its kinetic energy. But it is also clear that the energy
of the electron must be {\it on average} its kinetic energy, because
any other assertion requires too many additional assumptions to be
reconciled with the energy principle.

\section{Oscillating electron densities and
potentials}\label{wave-1}

Concerning the electron, one could ask what property of an electron
could actually be responsible for its wave features, if these are
taken to be real. The obvious answer to this question, and one which
will not seem strange to condensed matter theorists, is {\em
electron charge (or mass) density}. In condensed matter these charge
density waves are actually routinely observed, in particular in
atomic scale experiments on metal surfaces.

It is therefore also quite natural to assume that the wave
properties will be related to some form of density oscillation.
Here, we assume that the number density varies in the interval
[0,1], where 1 indicates the density maximum. In the simplest case,
that of a free electron traveling at a constant velocity $v_{el}$ in
$z$-direction, these oscillations are described by the plane wave:
\begin{eqnarray}
\rho(z,t) = \frac{\rho_{0}}{2}\left[1 + \cos\left(\frac{4 \pi}{\lambda} z - 4 \pi \nu t\right)\right] \nonumber \\
\end{eqnarray}
The amplitude $\rho_{0}$  will be subject to constraints, for
example the condition that the density integrated over a certain
volume equals unity. This condition, however, need not concern us at
this point and it will in fact be shown that all fundamental results
necessary can be derived without this normalization condition.
 It is important, though, to check that
the ansatz is compatible with the limit of inertia. For vanishing
velocity we get, using Eqs. (\ref{planck1}) and (\ref{broglie1}):
\begin{eqnarray}
\lim_{v_{el} \rightarrow 0} \rho = \rho_{0},
\end{eqnarray}
which is compatible with our basic assumption that the wave
properties of electrons and their oscillating density are a direct
consequence of their state of motion. An oscillating density of
charge for a free electron is ruled out, if no external or internal
potential energy components are present, since it violates the
energy conservation principle. We thus have to introduce potentials
to make up for the periodic variations. As the simplest case, these
could be thought of as oscillating $E$ or $B$ fields. The reason for
choosing vector fields rather than scalar ones lies in the one known
additional property of an electron: electrons possess an intrinsic
magnetic moment or a spin. Since such a property is incompatible
with a scalar distribution of field properties, it has to be related
in the simplest case at least to some kind of vector field. Whether
this is sufficient, has to be established subsequently by an
analysis of interactions of electrons with external electro-magnetic
fields. The two additional fields we introduce are an electric
$\cal{E}$ and a magnetic $\cal{H}$ field, which are thought
transverse as in the case of photons, but with double the wavelength
and consequently half the frequency of the density oscillations.
\begin{eqnarray}
{\cal{E}} = {\bf e}_1 {\cal{E}}_0 \cos\left(\frac{2 \pi}{\lambda} z
- 2 \pi \nu t +
\phi\right) \nonumber \\
{\cal{H}} = {\bf e}_2 {\cal{H}}_0 \cos\left(\frac{2 \pi}{\lambda} z
- 2 \pi \nu t + \phi\right)
\end{eqnarray}
The helicity in this way can either be positive - it complies with
the standard right-hand behaviour - or negative - it is the
opposite. This feature introduces two possible groundstates of the
free electron, which shall be later identified as its spin-up and
spin-down state. The additional phase $\phi$ has been added to
account for energy conservation of electrons at the local level.

Electromagnetic components to the electron's mass are not a new
concept. They have also been proposed by Abraham and Lorentz in
their ''classical'' models of electrons \cite{abraham03,lorentz04}.
Compared to these classical models, the present one is different in
two aspects: (i) The shape of the electron, e.g. a point or a
sphere, is not imposed from the outset. Such a shape would show up
in high-energy scattering events in the scattering cross sections.
While this is true for protons or, more generally, for atomic
nuclei, it has never been found for electrons. One reason could be
that electrons actually do not possess a defined shape, but that
their shape depends on the potential environment. (ii) The
electromagnetic components are not constant, but depend on the state
of motion of the electron. This is, as analyzed in previous
sections, necessary to reconcile the assumption with the wave
features of electrons.

Energy conservation requires that the energy density at every single
point of the electron is a constant. The energy density of the field
components at a given point is:
\begin{eqnarray}
E_{field} &=& \frac{1}{2}\epsilon_0{\cal{E}}^2 + \frac{1}{2} \mu_0
{\cal{H}}^2  \\ &=& \left(\frac{1}{2}\epsilon_0{\cal{E}}_0^2 +
\frac{1}{2} \mu_0 {\cal{H}}_0^2\right)\cos^2\left(\frac{2
\pi}{\lambda} z - 2 \pi \nu t + \phi\right)\nonumber
\end{eqnarray}
The simplest solution to the constant energy density problem, which
is also the simplest solution for the wave propagation of an
electron is a phase shift of the fields by $\pi/2$, so that
\begin{eqnarray}
&\phi& = \frac{\pi}{2} \\ &\Rightarrow& \quad E_{field} =
\left(\frac{1}{2}\epsilon_0{\cal{E}}_0^2 + \frac{1}{2} \mu_0
{\cal{H}}_0^2\right) \sin^2\left(\frac{2 \pi}{\lambda} z - 2 \pi \nu
t\right)\nonumber
\end{eqnarray}
At this stage the introduction of fields with the given frequencies
and wavelengths becomes clear, because we can eliminate periodic
components of the total energy density with the help of the relation
for the cosine at half angles:
\begin{equation}
2 \cos^2(x) = 1 + \cos(2 x)
\end{equation}
The kinetic energy density is then, in a first step:
\begin{equation}
E_{kin} = \frac{1}{4} \rho_{0} v_{el}^2 \left[1 + 2
\cos^2\left(\frac{2 \pi}{\lambda} z - 2 \pi \nu t\right) - 1\right]
\end{equation}
And with the following ansatz for the amplitudes of the fields:
\begin{equation}
\left(\frac{1}{2}\epsilon_0{\cal{E}}_0^2 + \frac{1}{2} \mu_0
{\cal{H}}_0^2\right)  = \frac{1}{2}\rho_{0} v_{el}^2
\end{equation}
We get for the total energy density:
\begin{eqnarray}
E_{tot} &=& \frac{1}{4} \rho_{0} v_{el}^2  + \frac{1}{4} \rho_{0}
v_{el}^2\left[2 \cos^2\left(\frac{2 \pi}{\lambda} z
 - 2 \pi \nu t\right) \right. \nonumber \\ &+& \left. 2 \sin^2
 \left(\frac{2 \pi}{\lambda} z - 2 \pi \nu t\right)- 1 \right]
= \frac{1}{2} \rho_{0} v_{el}^2
\end{eqnarray}
In contrast to the statement in Eq. (\ref{total_simple}), the model
leads to the result that the total energy is equal to the {\em
kinetic energy} of the electron, as also assumed in quantum
mechanics. At a fundamental level, this is due to one feature of the
ansatz: the wavelengths of field components and mass density
components are different. The electron's energy density is
determined by its mass density at the groundstate, i.e. $ v_{el}
\rightarrow 0$, {\em even though} the electron wave is a physically
real - i.e. with a physical property, the density, periodic in space
and time - feature. The relation between the frequency $\nu$ and the
wavelength $\lambda$ of this wave is the same as for a de Broglie
wave, i.e., the group velocity is equal to:
\begin{equation}
v_g = \frac{d \omega}{d k} = \frac{d \left(m v_{el}^2/2 \hbar
\right)}{d \left(m v_{el}/\hbar\right)} = v_{el}
\end{equation}
At this point the model captures at least four fundamental
properties of the electron in the conventional model. These
properties are:

\begin{itemize}
\item The wavelength of the electron wave is inverse proportional to its momentum (de Broglie).
\item The frequency of the electron wave is proportional to its kinetic energy (Planck).
\item The total energy of the electron is just the kinetic energy of its rest mass at every point of wave propagation (Energy conservation).
\item The total density of a free electron including the fields is a constant and equal to the inertial electron density (quantum mechanical plane wave).
\end{itemize}

So far all variables introduced, the density as well as the fields
are thought to be physical quantities, i.e. they should be in
principle measurable. But the ansatz also leads to the conclusion
that a wavefunction, if defined on the basis of this model and in
line with the required properties, {\em cannot be} a physical
quantity. This is shown in the following section.

\subsection{Stability of free electrons}

Before, we have to consider an additional problem, which becomes
imminent, as soon as the electron is considered to be a structure
with a finite extension in space and not a point-particle without
structure: the problem of Coulomb repulsion. In DFT this problem
arises due to the requirement that one electron does not interact
with itself. Exchange and correlation potentials thus have to be
corrected for this constraint \cite{perdew81}. The situation is
different in classical electrodynamics, where the electron, as a
point particle, would carry infinite electrostatic energy(see p. 751
of Ref. \cite{jackson-ed}). This has been the starting point of
early attempts by Abraham and Lorentz to circumvent the problem by
stating that the electron's mass was purely electromagnetic
\cite{abraham03,lorentz04}. The electron radius, its classical
radius, is then about 2.8~fm. However, this is still orders of
magnitude larger than the largest possible radius inferred from
high-energy scattering experiments. In relativistic quantum
electrodynamics the problem of electron self-energy is removed by
renormalization \cite{dela04}.

In the present context the problem is rather trivially solved (with,
of course, the remaining problem of electron self-interference, see
the discussion), by defining a {\em cohesive static potential}
$\phi_{coh}$, depending on the electron density. Electron charge
densities are typically one electron per sphere with a Wigner-Seitz
radius $r_s$ of about two atomic units, as found in solids. We start
from a spherical charge distribution of density $\rho_{0}$, with a
radius of $r_s$. Then the repulsion energy is given by the following
integral (we use atomic units in the following derivation) :
\begin{eqnarray}
W &=& \rho_{0}^2 \int_0^{r_s} dr \frac{4 \pi r^3}{3} \frac{1}{r} 4
\pi r^2 \nonumber \\&=& \left(\frac{ 4 \pi}{3} r_s^3 \rho_{0}\right)
\left(\frac{4 \pi}{3} r_s^3 \rho_{0}\right) \frac{3}{5r_s} =
\frac{3}{5 r_s}
\end{eqnarray}
The potential associated with this repulsion, the Hartree potential, is thus:
\begin{equation}
V_H = V_H(r_s) = \frac{3}{5} r_s^{-1}
\end{equation}
For electrons in metals the typical value for $r_s$ is about two
Bohr radii. There is no reason to assume that the density for a free
electron will be substantially different. Thus the additional {\em
cohesive potential} for a free electron should be:
\begin{equation}
V_{coh} = - \frac{3}{5 r_s} \approx (r_s = 2 au) \approx - 8.16 eV
\end{equation}
For a free electron the sum of both potentials must be zero.
The repulsive and attractive fields are
($\Omega$ is the volume of the Wigner-Seitz cell):
\begin{eqnarray}
\frac{W_{coh}}{\Omega} &=& - \rho_{0} \phi_{coh} = - \rho_{0}
\left(\frac{ 4 \pi}{3}
r_s^3 \rho_{0}\right) \frac{3}{5r_s} \nonumber \\
\phi_{coh} &=& - \frac{3}{5r_s} \qquad \phi_H  = \frac{3}{5r_s}\nonumber \\
\phi_{el,0} &=& \phi_{coh} + \phi_{H} = 0
\end{eqnarray}
Summarizing the results of this analysis we find that a model of
free electrons, which refers wave properties to oscillating charge
densities and electromagnetic potentials, can be in line with
conventional experimental and theoretical models because the total
energy of the electron is in both cases equal to the kinetic energy
of the electron, and complies with the fundamental Planck and de
Broglie relations. The repulsive Hartree potential within a single
electron is corrected by a cohesive potential, which depends on the
density of electron charge. The total electrostatic potential in
this case vanishes, as required by all conventional theoretical
models of single electrons. For the following derivations we use
atomic units, where $e=m=c=\hbar=1$.

\subsection{Why wavefunctions are not physical objects}

The requirements for a wavefunction $\psi$ or a spinor in the
context of the Dirac theory are that the wavefunction can be written
as a complex number (vector for spinors), that a duality operation
$\psi \rightarrow \psi^{\dagger}$ exists, and that the product of
$\psi$ and $\psi^{\dagger}$ is positive and equal to the number
density of the electron. In atomic units the number density is equal
to the mass density and also, multiplied with the square of the
velocity of light in vacuum, the energy density. In the following
analysis this eases the notation considerably, while it would have
to be complemented by suitable constants if performed in the SI
system of units. Given the two separate energy contributions in our
model, we can readily identify the real part of the wavefunction as
the square root of the number density, or $\rho^{1/2}$. The
imaginary part is more difficult to pin down. The requirement here
is that it must be related to the magnetic properties of the
electron, or its ''spin''. The simplest ansatz is due to geometric
algebra and the geometric product between vectors.
 Following our model we assume that the
velocity vector of the electron is parallel to ${\bf e}_3$. For ease
of notation we introduce here a Poynting-like vector, where the
electromagnetic energy flux is given by (in atomic units c = 1):
\begin{equation}
{\cal{S}} = {\cal{E}}{\cal{H}} = {\bf e}_1 {\bf e}_2 {\cal{E}}_0
{\cal{H}}_0 \sin^2\left(\frac{2 \pi}{\lambda} z - 2 \pi \nu t\right)
\end{equation}
Then the energy density of field components is given by the
following scalar:
\begin{eqnarray}
{\cal{S}} &=& i {\bf e}_3 S \\
S &=& {\cal{E}}_0 {\cal{H}}_0 \sin^2\left(\frac{2 \pi}{\lambda} z -
2 \pi \nu t\right)
\end{eqnarray}
\begin{equation}
S = S_0 \sin^2\left(\frac{2 \pi}{\lambda} z - 2 \pi \nu t\right)
\qquad S_0 \equiv {\cal{E}}_0 {\cal{H}}_0
\end{equation}
If we require that the real part of the wavefunction does not change
under a transformation from positive to negative helicity, and that
the imaginary part is antisymmetric, then we can write the
wavefunction in the following way:
\begin{equation}
\psi = \rho^{1/2} +  S^{1/2}{\bf e}_1{\bf e}_2 = \rho^{1/2} +  i
S^{1/2} {\bf e}_3
\end{equation}
The wavefunction in this case is a multivector composed of a scalar
(the ''real'' component) and a bivector (the ''imaginary''
component). The duality operation changes the helicity of the
electron; so exchanging the direction of ${\cal{E}}$ and ${\cal{H}}$
we obtain:
\begin{equation}
\psi^{\dagger} = \rho^{1/2} +  S^{1/2} {\bf e}_2{\bf e}_1 =
\rho^{1/2} -  i S^{1/2} {\bf e}_3
\end{equation}
For the product of $\psi$ and $\psi^{\dagger}$ we get consequently:
\begin{equation}
\psi^{\dagger}\psi = \psi \psi^{\dagger} = \rho + S
\end{equation}
And if we set, as before:
\begin{eqnarray}\label{rho0}
\rho &=& \rho_0 \cos^2\left(\frac{2 \pi}{\lambda} z - 2 \pi \nu t \right) \nonumber \\
S &=& S_0 \sin^2 \left(\frac{2 \pi}{\lambda} z - 2 \pi \nu t\right) \nonumber \\
S_0 &=& \rho_0
\end{eqnarray}
Then we obtain the result that the product $\psi^{\dagger}\psi$
corresponds to the inertial number density of the electron:
\begin{equation}
\psi^{\dagger}\psi = \rho + S = \rho_0 = constant
\end{equation}
which is the standard result in quantum mechanics and also in
density functional theory. However, within the present framework
wavefunctions also have a physical content, contrary to their role
in the standard model where only their square has a meaning, that of
a probability density. One could pin down the difference by saying
that even though the wavefunction is certainly not a physical object
of the same reality as an electromagnetic field vector or a scalar
potential, it contains physically real objects like fields and mass
(or charge) densities - or rather the roots thereof.

The multivector $\psi$ does not comply with the time-dependent
Schr\"odinger equation, because the single time derivative of $\psi$
will yield the sum of a vector and an imaginary number (the ''odd''
elements of a multivector in three dimensions). In our view this
bears on the fact that the model developed so far is not covariant.
For applications in DFT this is not relevant, as the LPS Eq.
(\ref{lps-01}) is based on the time-independent Schr\"odinger
equation. In order to compare the extended model of the electron to
standard theory, we therefore define the Schr\"odinger wavefunction
$\psi_S$ as a complex number, retaining the direction perpendicular
to the field vectors ${\bf e}_3$ as a hidden variable:
\begin{eqnarray}
\psi_S &\equiv& \rho_0^{1/2} \left[\cos\left(\frac{2 \pi}{\lambda} z
- 2 \pi \nu t \right) + i \sin\left(\frac{2 \pi}{\lambda} z - 2 \pi
\nu
t \right)\right]\nonumber \\
&=& \rho_{0}^{1/2} \exp i \left(\frac{2 \pi}{\lambda} z - 2 \pi \nu
t \right)
\end{eqnarray}
The kinetic energy operator or the time differential acting on
$\psi_S$ then returns the energy eigenvalue of the free electron:
\begin{eqnarray}
- \frac{1}{2}\nabla^2 \psi_S &=& - i^2 \frac{1}{2} \frac{4
\pi^2}{\lambda^2}
\psi_S = \frac{1}{2} v_{el}^2 \psi_S = E \psi_S \nonumber \\
i \frac{\partial}{\partial t} \psi_S &=& -i^2 \omega \psi_S = \omega
\psi_S = E \psi_S
\end{eqnarray}
It is interesting to note that the current density ${\bf J}$, as
defined from the continuity equation and the time-dependent
Schr\"odinger equation will be a function without local density
variations. From the continuity equation
\begin{eqnarray}
\frac{\partial \rho}{\partial t} + \nabla \cdot {\bf J} =
\frac{\partial}{\partial t} \psi_S^{\dagger} \psi_S + \nabla \cdot
{\bf J} = 0,
\end{eqnarray}
we obtain the following result for ${\bf J}$:
\begin{eqnarray}
{\bf J} &=& \frac{1}{2i} \left(\psi_S^{\dagger} \nabla \psi_S - \psi_S \nabla \psi_S^{\dagger}\right) \\
{\bf J} &=& \frac{2 i}{2 i} \left[(\rho)^{1/2}\nabla S^{1/2} -
S^{1/2}\nabla (\rho)^{1/2}\right]
\nonumber \\
        &=& \rho_0\left[ \cos\left(\frac{2 \pi}{\lambda} z - 2 \pi \nu t\right)\frac{\partial}{\partial z}
        \sin\left(\frac{2 \pi}{\lambda} z - 2 \pi \nu t\right)\right.
        \nonumber \\
        &-& \left. \sin\left(\frac{2 \pi}{\lambda}z - 2 \pi \nu t\right)\frac{\partial}{\partial z}
        \cos\left(\frac{2 \pi}{\lambda}z - 2 \pi \nu t\right) \right]
        {\bf e}_3 \nonumber \\
        &=& \rho_0 v_{el} \left[sin^2 \left(\frac{2 \pi}{\lambda}z - 2 \pi \nu t\right)
        + cos^2 \left(\frac{2 \pi}{\lambda}z - 2 \pi \nu t\right)\right] {\bf e}_3
        \nonumber \\
        &=& \rho_0 v_{el} \, {\bf e}_3 \nonumber
\end{eqnarray}
which is equal to the momentum density of the inertial electron
mass. It thus gives a consistent picture of the electron as an
entity without any inner structure.

To summarize the findings of this analysis we may say that even
though this model electron is an entirely physical object, all of
whose properties are described be standard physical quantities, we
can only describe it as a scalar complex number, i.e. a
wavefunction, if this wavefunction itself does not have a direct
physical meaning.

\section{Electrons in static external fields}\label{static}

While quantum mechanics, as shown in the previous section, describes
an electron by a complex number, a wavefunction, which is not
strictly speaking a physical object, it does so consistently by
assigning a current density to the propagation of the electron's
inertial mass. Then, of course, an external electric or magnetic
field cannot affect the density distribution of the electron in a
physically transparent manner. Within the present context, where all
components of an electron in motion are actually physical objects,
we have to develop an understanding, at a very fundamental level,
what actually takes place if an electron is accelerated. Let us
assume that this is due to an external potential $\phi$. From a
physical point of view, four discrete processes are bound to happen:

\begin{enumerate}
\item The electron velocity will change.
\item The electron's density distribution will change.
\item The electromagnetic field components will change.
\item The intensity of the external field $\phi$ will be diminished due
to energy transfer.
\end{enumerate}

The way to account for all four processes is captured in the
following equation:
\begin{equation}\label{newton1}
{\bf f} = - \nabla \phi = \rho_0 \frac{d {\bf v}_{el}}{d t}
\end{equation}
The sink in the scalar field $\phi$ here accounts for the transfer
of energy from the external field to the electron. It is easy to
prove that the internal variations of the mass density or the field
$S$ do not change this equation. We know from the preceding
derivations that
\begin{equation}
\rho + S =  \rho_0 = constant
\end{equation}
Then it follows that:
\begin{equation}
\dot{S} + \dot{\rho} = 0 \quad \rightarrow \quad
\frac{d}{dt}(\psi_S^{\dagger}\psi_S) v = \rho_0 \frac{d v_{el}}{dt}
\end{equation}
Therefore Eq. (\ref{newton1}) is valid without restrictions. Again,
as for the current density, we find that the density variations are
hidden in the combined effect on density and field contributions.
Interestingly, this is only the case because the time derivative of
the density variations is of equal magnitude but opposite sign as
the derivative of the field variations.

\subsection{The problem of frequency}

At this point it is unclear, what the frequency of an electron wave
after acceleration in an external potential will be. We cannot use
the conventional reasoning in quantum mechanics that the total
energy is reflected by the frequency, while the wavelength is
connected to the kinetic energy. This reasoning, which is one of the
heuristic arguments to introduce the Schr\"odinger equation, is only
justified on the basis of classical mechanics. In particular, it is
still unclear, how an external potential actually influences the
frequency of an electron wave. However, a closer look at the
fundamental {\em experimental} results leading to the development of
quantum mechanics reveals that the change of frequency with the
intensity of electrostatic fields is already implicitly contained in
Einstein's work on photoelectron emission. There, the kinetic energy
of a metal electron and its frequency are related to the frequency
of the incident photon and the intensity of the metal potential
$\phi_m$ via:
\begin{equation}
\frac{v_{el}^2}{2} = \omega_{el} = \omega_{ph} - \phi_m
\end{equation}
Generalizing this result, it can be said that the frequency
$\omega_{el}$ of an electron in an electrostatic field, compared to
the frequency $\omega_{el}^0$ of a free electron, shifts due to the
existence of an electrostatic potential with:
\begin{equation}
\omega_{el}(\phi_m) = \omega_{el}^0 - \phi_m
\end{equation}
Then the density components and the field components of the electron in an external potential $\phi$ will comply with:
\begin{eqnarray}
\psi_S &=& \rho^{1/2} + iS^{1/2} \nonumber \\
\rho^{1/2} &=& \rho_0^{1/2} \cos\left[\frac{2 \pi}{\lambda} z - (\omega_0 - \phi_m) t \right] \nonumber \\
S^{1/2} &=& \rho_0^{1/2} \sin\left[\frac{2 \pi}{\lambda} z - (\omega_0 - \phi_m)  t \right] \nonumber \\
\psi_S &=& \rho_0^{1/2} \exp i\left[\frac{2 \pi}{\lambda} z -
(\omega_0 - \phi_m) t \right]
\end{eqnarray}
It should be clearly understood at this point that the frequency
transferred to the electron is not the original frequency of the
photon, but that this frequency is diminished, within the metal, due
to the effect of the metal's electrostatic field. This diminished
frequency is the frequency which the photon could actually transfer
to the electron. The free electron, which is then measured, has this
diminished frequency and a wavelength, which corresponds to this
frequency: a mechanism of enforcing a wavelength, which we shall
analyze in some details further down when we present a detailed
model of photon-electron interactions. At this point it becomes
clear that this wavefunction, corrected for a change of frequency
due to electrostatic potentials compared to the wavefunction of a
free electron, complies with the general Schr\"odinger equation for
electrons in a potential $V = - \phi_m$:
\begin{eqnarray}\label{schro}
- \frac{1}{2} \nabla^2 \psi_S &=& \frac{v_{el}^2}{2} \psi_S = \omega_0 \psi_S  \nonumber \\
i \frac{\partial}{\partial t} \psi_S &=& (\omega_0 - \phi_m) \psi_S \nonumber \\
i \frac{\partial \psi_S}{\partial t} &=& \left[ - \frac{1}{2}
\nabla^2 + V \right] \psi_S
 \end{eqnarray}
Again, it should be kept in mind that while the wavefunction itself
is not a physical object, but a multivector in geometric algebra,
its components, the real part and the bivector or imaginary part,
are related to real physical properties. Here, we have assumed one
direction of motion of the electron. It is, however, clear that the
same general relation must apply to a general direction of motion.
Eq. (\ref{schro}) is therefore generally valid.

Considering, for example, an electron in a one-dimensional well $V$
of length $L$, we find the usual result that $k = 2 \pi /L \cdot n$.
In this case a superposition of left and right traveling waves
corresponds to:
\begin{eqnarray}
\psi^{+} &=& \rho^{1/2} + i {\bf e}_3 S^{1/2} \nonumber \\
\psi^{-} &=& \rho^{1/2} - i {\bf e}_3 S^{1/2} \nonumber \\
\psi &=& \frac{1}{2} \left(\psi^{+} + \psi^{-}\right) = \rho^{1/2} \nonumber \\
\psi^2 &=&  \rho
\end{eqnarray}
The density distribution measured, e.g. in Mike Crommie's quantum
corral experiments \cite{eigler93}, is then the {\em real} electron
density and not the probability density.

\subsection{Electron in a magnetic field}

The acceleration of electrons in an external magnetic field is, conventionally,
described by the Lorentz force equation:
\begin{equation}
\rho_0 \frac{d {\bf v}}{d t} = \rho_0 \left({\bf E} + {\bf v} \times
{\bf B}\right)
\end{equation}
In geometric algebra, the same relation is expressed in a Lorentz
covariant manner and using the Faraday multivector F (see p. 157 of
Ref. \cite{dor02}):
\begin{equation}
F = {\bf E} + i {\bf B} \qquad \dot{v} = F \cdot v
\end{equation}
We have shown in the previous sections dealing with electron
acceleration that the electron's momentum change is described by the
product of the inertial electron density times the change of the
electron velocity. This result is in accordance with the standard
model, e.g. in density functional theory, where the electron does not possess an intrinsic
and variable structure. Here, we only need to add that the same
applies to the component of the Lorentz force which is perpendicular
to the electron's velocity. A separate verification  is therefore
not required.

The situation becomes more interesting, if we analyze the change of
the electron's intrinsic structure in a constrained trajectory. One
can think, for example, of an electron contained in the potential
well of a atomic nucleus, which has to follow the velocity vector of
the nucleus. Then, the changes due to the external magnetic field
must act on the electron's intrinsic properties as well. The
important question, in the context of quantum mechanics, is how an
external field will affect the electron's ''spin''. However, we have
not yet defined what the spin of an electron actually is in the
present model. Following Doran and Lasenby \cite{dor02} we could
define the spin of an electron as:
\begin{equation}
{\bf s} = \frac{1}{2} \psi {\bf e}_3 \psi^{\dagger}
\end{equation}
Since $\psi = \rho^{1/2} + {\bf e}_1 {\bf e}_2 S^{1/2}$ and
$\psi^{\dagger} = \rho^{1/2} - {\bf e}_1 {\bf e}_2 S^{1/2}$ the spin
in this case is perpendicular to the field plane:
\begin{eqnarray}
\psi {\bf e}_3 \psi^{\dagger} &=& \rho {\bf e}_3 - (\rho S)^{1/2}{\bf e}_3{\bf e}_1{\bf e}_2 +
(\rho S)^{1/2}{\bf e}_1{\bf e}_2{\bf e}_3 \nonumber \\ &-& S {\bf e}_1{\bf e}_2{\bf e}_3{\bf e}_1{\bf e}_2 \nonumber \\
&=& (\rho + S){\bf e}_3 + (\rho S)^{1/2}\left({\bf e}_1{\bf e}_2 + {\bf e}_2{\bf e}_1\right){\bf e}_3 = \rho_0 {\bf e}_3 \nonumber \\
{\bf s} &=& \frac{1}{2} \rho_0 {\bf e}_3
\end{eqnarray}
The spin, defined in this manner, is a constant vector associated
with the direction perpendicular to the $S$ plane. However, this is
not compatible with the experimental results on Stern-Gerlach
experiments \cite{stern22}, because an external magnetic field
perpendicular to the electron's trajectory would lead to a vanishing
magnetic moment. Calculating the equivalent expression for the
direction ${\bf e}_2$, which is perpendicular to the direction of
motion, we obtain:
\begin{eqnarray}
\psi {\bf e}_2 \psi^{\dagger} &=& \left(\rho^{1/2} + i {\bf e}_3
S^{1/2} \right){\bf e}_2 \left(\rho^{1/2} - i {\bf e}_3 S^{1/2}
\right) \nonumber \\
&=& \rho^{1/2} {\bf e}_2 \rho^{1/2} - \rho^{1/2}{\bf e}_2 i {\bf
e}_3 S^{1/2} + i {\bf e}_3 S^{1/2}{\bf e}_2 \rho^{1/2} \nonumber \\
&-& i {\bf e}_3 S^{1/2} {\bf e}_2 i {\bf e}_3 S^{1/2}
 \nonumber \\
&=& (\rho + S){\bf e}_2 + i (\rho S)^{1/2}\left({\bf e}_3{\bf e}_2 -
{\bf e}_2{\bf e}_3\right) \\
&=& \rho_0 \left[{\bf e}_2 + \sin\left(\frac{4 \pi}{\lambda} z -
2\omega t \right) {\bf e}_1\right]\nonumber
\end{eqnarray}
In the last lines we have used the anticommutation of frame vectors
and the trigonometric relations for half angles. The directions
${\bf e}_1$ and ${\bf e}_2$, perpendicular to the direction of
electron motion, yield equivalent results: a constant vector in one
direction and an oscillating vector in the direction perpendicular
to it. The spin vector, defined in the same manner as above, would
then be:
\begin{equation}
{\bf s} = \frac{1}{2}\psi {\bf e}_2 \psi^{\dagger} =
\frac{\rho_0}{2} \left[{\bf e}_2 + \sin\left(\frac{4 \pi}{\lambda} z
- 2\omega t \right) {\bf e}_1\right]
\end{equation}
Here we slightly modify our electron model: we assume that the plane
of electromagnetic fields is not perpendicular to the velocity
vector, but that  the velocity vector encloses an angle of
45$^{\circ}$ with its projection in the plane. The spin vector for
this electron can then be described by:
\begin{equation}
{\bf s}^{\pm} = \frac{1}{2} \rho_0 \left[\pm \frac{{\bf e}_3 + {\bf
e}_2}{\sqrt{2}} \pm \frac{1}{\sqrt{2}}\sin\left(\frac{4
\pi}{\lambda} z - 2\omega t \right) {\bf e}_1\right]
\end{equation}
The spin average over one period $\tau = 1/\nu$ will then be:
\begin{eqnarray}
\langle{\bf s}^{\pm}\rangle = \frac{1}{\tau} \int_{0}^{\tau} {\bf
s}^{\pm} dt  = \pm \frac{1}{2} \rho_0 \frac{{\bf e}_3 + {\bf
e}_2}{\sqrt{2}}
\end{eqnarray}
It should be clear, though, that the previous derivations remain
valid also for this modification of the electromagnetic plane. The
reason for this is that the scalar and bivector components are
separately accounted for and every cross product will vanish for a
term $\psi^{\dagger} \psi$. The orientation of the bivector plane is
a free variable in the description, the energy conservation
principle restricts only the phase of the wave, but not the
orientation of the plane. According to postulate 3 of the
introduction the magnetic properties of the electron should be due
to the orientation of the fields, which in the previous equation has
been linked to the orientation of the spin vector. Then we may
define a magnetic moment density of the electron by:
\begin{equation}
\vec{\mu} = \gamma \langle{\bf s}\rangle = \gamma \frac{\rho_0}{2}
\frac{\pm ({\bf e}_3 + {\bf e}_2)}{\sqrt{2}}
\end{equation}
In an external magnetic field, aligned for example in the ${\bf
e}_2$ direction, the magnetic moment density will interact with the
field and lead to a variation of the static potential according to:
\begin{equation}
\phi_{B} = - \vec{\mu} \cdot {\bf B}_{ext} = - \vec{\mu} \cdot B_0
{\bf e}_2
\end{equation}
This static potential will affect the frequency of the electron wave
in the same way as an electrostatic potential. Writing, for
efficiency, the wavefunction of the system in the magnetic field as:
\begin{equation}
\psi = \psi_0 e^{- i \omega_B t}
\end{equation}
where $\psi_0$ encapsulated all other degrees of freedom of the
electron wave and it is assumed that we can completely separate the
magnetic degrees of freedom in the wavefunction $\psi$. Then from
the time-dependent Schr\"odinger equation we get immediately:
\begin{eqnarray}
i \frac{\partial \psi}{\partial t} &=& \omega_B \psi = \pm
\frac{\gamma B_0}{2} \frac{\rho_0}{\sqrt{2}} \psi \nonumber \\
\omega_B &=& \pm \frac{\gamma B_0}{2} \frac{\rho_0}{\sqrt{2}}
\end{eqnarray}
Within the framework of geometric algebra and using rotors instead
of complex exponentials the wavefunction $\psi$ rotates around the
direction of the external $B$ field. The resulting effect
is much the same: the interaction of the field-components with the
external field leads to a rotation of the spin vector around the
magnetic axis with two opposite vectors of magnetic moment
\cite{challin96}. In a Stern-Gerlach experiment, in this case, the
ensuing trajectory will either be deflected along the positive or
the negative $y$ axis: exactly the result observed in the actual
experiments \cite{stern22}.

\subsection{Spin isotropy}

In standard theory, the spin of electrons is isotropic. This means,
that the direction of the spin vector without an external magnetic
field is undefined. It seems at first glance that the previous
section, where the spin vector was oriented in a definite direction
with respect to the vector of electron propagation, contradicts the
model in standard theory. If, for example, the external magnetic
field ${\bf B}$ is oriented along the ${\bf e}_1$ axis, then the
static potential $\phi_B$ will be zero. Clearly, this contradicts
the observations in Stern-Gerlach type experiments, where the split
into two separate trajectories is independent of the orientation of
the field.

The aim of this section is to show, how the concept of isotropic
spin fits into the presented model of electrons. From preceding
sections we know that electrons with vanishing velocity do not
possess field components to their energy density. If the plane of
these field components is related to the direction of spin, then
electrons of zero velocity do not possess spin.

As the electron only possesses spin, if it is in motion, the
question arises, how the spin vector relates to the plane of the
electromagnetic energy components. The electric ${\cal E}$ and
magnetic ${\cal H}$ components are in the plane perpendicular to the
velocity vector of the electron, the vector ${\cal E} \times {\cal
H}$ can therefore either be parallel (positive helicity) or
anti-parallel (negative helicity) to the velocity vector. Then the
spin vector of a free electron is described by the following
relations:
\begin{equation}
{\bf s}_0^\pm = \pm \frac{1}{2} \psi {\bf e}_3 \psi^\dagger = \pm
\frac{1}{2} \rho_0 {\bf e}_3
\end{equation}
In this case the spin--vector of a single electron is isotropic with
respect to the plane perpendicular to the direction of motion. As
this direction is singled out due to the motion of the electron, the
spin of a {\em single} electron cannot be completely isotropic, if
spin is to be a consequence of electron motion. However, for a
statistical manifold of $n^+$ spin--up (${\bf s}_0^+$) and $n^-$
spin--down (${\bf s}_0^-$) electrons, where $n^+ = n^-$ the manifold
will be fully isotropic in 3-dimensional space. So far, the model is
thus in accordance with experimental evidence.

Considering the experimental evidence further, it is inconceivable
that the plane of the electromagnetic field components has a defined
orientation, which is not perpendicular to the vector of velocity,
{\em without} any external ${\bf B}$ field, as this would lead not
to two discrete experimental results but, depending on the
orientation of the field, to more or less continuous results.
However, it is not in contradiction with experiments, if the
orientation is well defined and not perpendicular to the velocity
direction {\em within} an external ${\bf B}$ field.

This analysis shifts the focus on the process considerably, because
in this case the key problem is not to find an orientation of the
spin vector, which yields two discrete results in every possible
measurements in an external gradient of ${\bf B}$ (which, of course,
will lead to the Pauli equation and Pauli matrices
\cite{goudsmit25,pauli27}), but to determine, how an external field
${\bf B}$ changes the orientation of the field plane. If the
external field {\em rotates} the field plane by an angle which is
proportional to the angle between the external field vector ${\bf
B}$ and the vector of electron propagation ${\bf v}_{el}$, then the
result is exactly the result obtained in the previous section: the
magnetic moment then is either positive or negative with respect to
the plane perpendicular to the external field and the possible
measurements are either a deviation in positive or in negative
direction, as observed in a Stern-Gerlach experiment. We have
assumed, in the previous section, that this angle is half the angle
between the field vector and the velocity vector. This choice,
however, is not compulsory. As the exact angle will have to be
determined by a careful analysis of experimental results, it is well
beyond the scope of this paper, which only seeks to establish, that
experimental results in static magnetic fields are in accordance
with the proposed model of extended electrons. In fact, as shown in the following,
this angle is contained in a constant, which describes the rotation of the field
components by a modified Landau-Lifshitz equation.

\subsection{Spin dynamics}

\begin{figure}
\centering
\includegraphics[width=7cm]{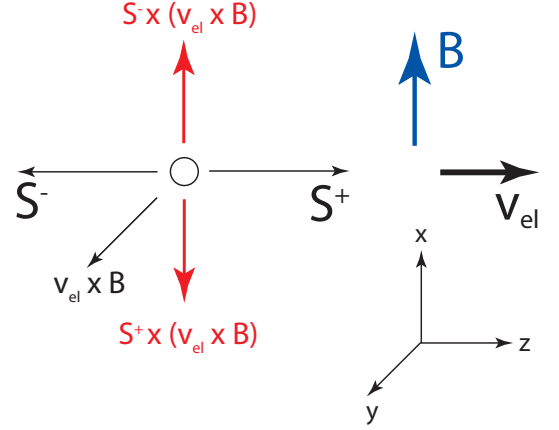}
\caption{Rotation of the field vector of an electron described by a modified Landau-Lifshitz equation.}
\label{Figure-spin}
\end{figure}

In condensed matter theory the dynamics of a spin system is described by the semi-empirical
Landau-Lifshitz equation, which reads \cite{landaulifshitz35}:
\begin{equation}
\frac{\partial {\bf M}}{\partial t} = - \frac{\gamma}{1 + \alpha^2} {\bf M} \times {\bf H}
- \frac{\gamma \alpha}{(1+\alpha^2) M_S} {\bf M} \times ({\bf M} \times {\bf H})
\end{equation}
In this equation ${\bf M}$ is the magnetization vector, ${\bf H}$ the applied magnetic field,
$M_S$ the saturation magnetization, and $\alpha$ and $\gamma$ an empirical constant and the gyromagnetic
ratio of the electron. As shown above, the field component ${\bf S}$ of the electron in motion will
precess around the magnetic vector if it is not parallel or antiparallel to the velocity. This is
described by the first part of the equation. The second part is commonly associated with
damping. Here, we propose a formulation, which is similar to the second component of the equation, but
 will lead to two discrete and anti-parallel induced spin directions. It includes the velocity vector
of the electron to describe something akin to a torque acting on the field vector. The change of direction
due to an external ${\bf B}$ field is described by:
\begin{equation}
\frac{d {\bf S}}{d t} = \mbox{const} \cdot {\bf S} \times \left({\bf v} \times \frac{d {\bf B}}{d t}\right)
\end{equation}
If the field is switched on in a finite interval, and to first order in the approximation, the induced field
vector within an external field is then described by:
\begin{equation}
{\bf S}_{induced} = \mbox{const} \cdot {\bf S} \times \left({\bf v} \times {\bf B}\right)
\end{equation}
 The equation has two discrete solutions (see Figure \ref{Figure-spin}). Given now that the field vector precesses around the external field we obtain exactly two solutions with opposite sign, which, in a field gradient along the $x$-direction will lead to two deflections in a Stern-Gerlach experiment.

We conclude the presentation of a model of electron interactions
with magnetic fields by pointing out that the key difference to
standard models is that spin isotropy is taken to its logical
extreme: the spin of an electron is only defined with respect to the
velocity vector of the electron, but not with respect to the
external magnetic field. External fields lead, as in standard
condensed matter physics, to a breaking of rotational symmetry. We
may thus conclude that the model of electrons developed accounts for
all known effects in static magnetic or electrostatic fields.

\subsection{Rotations and wavefunction symmetry}

The distinguishing property of fermions is that their wavefunction
changes its sign upon a rotation by $2 \pi$. To see that this is
also the case for our wavefunctions, we study a rotation of the spin
vector ${\bf s}$ of the free electron. In geometric  algebra a
rotation of a vector is accomplished by a two-sided multiplication
with a rotator $R$ (Ref. \cite{dor02}, p. 274):
\begin{equation}
{\bf s}' = R {\bf s} R^{\dagger}
\end{equation}
The one-sided rotator $R$ for a rotation of $\theta$ can be
described as the exponent:
\begin{equation}
R = R(\theta) = \exp - B \theta/2 \qquad B^2 = -1
\end{equation}
Since the spin vector is given by:
\begin{equation}
{\bf s} = \rho_0 \frac{{\bf e}^{\pm}}{2} = \psi \frac{{\bf
e}^{\pm}}{2} \psi^{\dagger},
\end{equation}
the wavefunction must transform in the following way:
\begin{eqnarray}
{\bf s}' &=& R(\theta) \psi \left[\frac{{\bf e}^{\pm}}{2}\right]
\psi^{\dagger} R^{\dagger}(\theta) \nonumber \\
\Rightarrow \quad \psi' &=& R(\theta) \psi
\end{eqnarray}
That the wavefunction transforms via a one-sided and not a two-
sided rotation is crucial for its symmetry properties. Assume now
that the angle of rotation $\theta = 2 \pi$. Then the spin vector
will make a full rotation and revert to its original state. But this
is not true for the wavefunction, since:
\begin{equation}
\psi' = \exp - B \pi \psi = \left[\cos (\pi) - B sin (\pi)\right]
\psi = - \psi
\end{equation}
One could thus conclude that also the general symmetry of the
wavefunction, which in many-electron systems is the basis of its
construction from Slater determinants, comes from geometrical
properties of the wavefunction as a sum of scalar and bivector
terms. We shall use this property in our construction of the
many-electron wavefunction.

\subsection{Vector potential and wavelength}

An interesting variation of the same topic is the behavior of
electrons in an Aharonov-Bohm experiment. There, the magnetic field
${\bf B}$ is actually zero, while the vector potential ${\bf A}$ is
non-zero. As measured orginally by Chambers \cite{chambers60} in 1960, and more
recently by Osakabe et al. in 1986 \cite{osa1986}, the
vector potential affects the phase of the electron. The effect can be formalised as
a change of the wavevector in an external field ${\bf A}$ with:
\begin{equation}
{\bf k}({\bf A}) = {\bf k}_0 + \alpha {\bf A}
\end{equation}
Here, $\alpha$ is assumed to be a constant. That the potential must
affect the wavevector or the local distribution of the electron
density and the related potentials can be inferred from the
description of a free electron after it has passed a region with
non-vanishing potential ${\bf A}$. In this case the wavefunction
$\psi_S(z,t)$ corresponds to the original wavefunction
$\psi_{S0}(z,t)$ augmented by an additional, and ${\bf A}$ dependent
phase. For simplicity we assume that ${\bf A}$ is parallel to ${\bf
e}_3$, then we get:
\begin{equation}
\psi_S({\bf A}, z, t) = \psi_{S0}(z,t) \cdot \exp \left(i \alpha
\int_{z_0}^{z} A(z) dz \right)
\end{equation}
Here, we have assumed that $A$ is only non-zero from z$_0$. Using
the formulation of $\psi_{S0}$ for a free electron derived above, we
may write for the wavefunction $\psi_S$ (z$_0$ is set to zero for
convenience):
\begin{equation}
\psi_S(A, z, t) = \exp i\left[ \left( \frac{2 \pi}{\lambda_0} +
\alpha A\right) z - 2 \pi \nu_0 t\right]
\end{equation}
Clearly the wavelength has been changed. If $A$ is positive then the
energy density in the region of the vector potential has been
increased. If this affects the electron, then it must lead to an
increase of the wavevector. The constant $\alpha$ must therefore be
positive. In atomic units and comparing with the experimental phase
shifts one finds 1. However, since the electron charge is negative
we get for the wavefunction $\psi$ in an external vector potential
the result:
\begin{equation}
\psi_S(A, z, t) = \exp i\left[ \left( \frac{2 \pi}{\lambda_0} -
A\right) z - 2 \pi \nu_0 t\right]
\end{equation}
The density and field components of the electron in this case are described by:
\begin{eqnarray}
\rho^{1/2} &=& \rho_0^{1/2} \cos\left[\left(\frac{2 \pi}{\lambda_0} - A\right) z - 2 \pi \nu_0 t \right] \nonumber \\
S^{1/2} &=& \rho_0^{1/2} \sin\left[\left(\frac{2 \pi}{\lambda_0} -
A\right) z - 2 \pi \nu_0 t \right]
\end{eqnarray}
In an Aharonov-Bohm experiment with two different pathways
characterized by opposite values of $A$ within the interval $z_0$,
$z_1$, the two partial waves at a point $z$ will be:
\begin{eqnarray}
\psi_S^+(A, z, t) = \rho^{1/2} \exp i \left(\frac{2 \pi}{\lambda_0}
z - (z_1 - z_0) A - 2\pi \nu t\right) \nonumber \\
\psi_S^-(A, z, t) = \rho^{1/2} \exp i \left(\frac{2 \pi}{\lambda_0}
z + (z_1 - z_0) A - 2\pi \nu t\right) \nonumber \\
\end{eqnarray}
A superposition of the two partial waves at $z$ will consequently
lead to an oscillating amplitude, described by:
\begin{eqnarray}
\psi_S &=& \frac{1}{2} \left(\psi^+ + \psi^-\right) \nonumber \\
&=& \rho^{1/2} \exp i \left(\frac{2 \pi}{\lambda_0}
z - 2\pi \nu t\right) \cos (z_1 - z_0) A \nonumber \\
&=& \left(\rho^{1/2} + i S^{1/2}\right) \cos (z_1 - z_0) A
\end{eqnarray}
This oscillation can then be measured, as in the case of electron
density waves at surfaces, as an $A$ and $z_1 - z_0$ dependent
density variation at a detector screen.

A generalisation of this result into a Schr\"odinger equation which
includes an external vector potential is not as straightforward as
in the case of electrostatic fields. The reason is that ${\bf A}$ is
not generally a vector, but a bivector, properly written as $i {\bf
A}$ (this follows from the fact that in a relativistic framework it
has to be a complex field \cite{dor02}). We start by writing the
wavefunction in a more transparent form as:
\begin{equation}
\psi_S = \rho_0^{1/2} \left[ \cos \left(({\bf k} - i{\bf A})\cdot
{\bf x} - \omega t\right) + i \sin \left( ({\bf k} - i {\bf A})
\cdot {\bf x} - \omega t\right) \right]
\end{equation}
Here, ${\bf x}$ is a three-dimensional position vector, and $i$ is
the pseudoscalar ${\bf e}_1{\bf e}_2{\bf e}_3$. If we assume that
the vector potential in the region of interest is constant, then a
first derivative of this wavefunction leads to the following result:
\begin{eqnarray}
\nabla \psi_S &=& \rho_0^{1/2} \left\{ - ({\bf k} - i {\bf A}) \sin
\left(({\bf k} - i {\bf A}) \cdot {\bf x} - \omega t\right) \right.
\nonumber \\
&+&  \left. (i {\bf k} + {\bf A}) \cos \left(({\bf k} - i {\bf A})
\cdot {\bf x} - \omega t\right) \right\}  \\
&=& \rho_0^{1/2}\left[ - {\bf k} \sin \phi + {\bf A} \cos \phi + i
({\bf k} \cos \phi + {\bf A} \sin \phi)\right] \nonumber
\end{eqnarray}
For brevity we have written $\phi$ for the argument of the periodic
functions in the last line. It can clearly be seen that the
inclusion of a general vector potential into the Schr\"odinger
equation requires that it has a different nature than the terms
related to the wavevector ${\bf k}$. The gradient of $\psi_S$ in
this case is a multivector composed of even (the bivector terms) and
odd (the vector terms) elements. For the second derivative one
obtains with the same procedure:
\begin{equation}
\nabla^2 \psi_S = \rho_0^{1/2} \left( -{\bf k}^2 - {\bf A} \cdot
{\bf k} - {\bf k} \cdot {\bf A} + {\bf A}^2\right) \psi_S
\end{equation}
To account for the changes in the local derivatives compared to the
frequency in a Schr\"odinger-type equation one consequently has to
set:
\begin{equation}
i \frac{\partial \psi_S}{\partial t} = \frac{1}{2} \left(i \nabla -
{\bf A}\right)^2 \psi_S
\end{equation}
Including also a scalar potential $V$, we obtain:
\begin{equation}
i \frac{\partial \psi_S}{\partial t} = \frac{1}{2} \left({\bf e}_i
\frac{i \partial}{\partial x_i} - {\bf A}\right)^2 \psi_S + V \psi_S
\end{equation}
The equation is equal to the general non-relativistic Schr\"odinger
equation for electrons in the presence of electrostatic and vector
potentials. The suggested extended model of electrons is thus fully
compatible with the non-relativistic theoretical framework of
quantum mechanics.

\section{Electrons in dynamic external fields}\label{dynamic}

Based on these findings one may equally analyze the effect of an
external electromagnetic field on the velocity of electrons. The
electromagnetic field shall be described by a bivector $i
S_{em}^{1/2}{\bf e}_3$, which propagates in ${\bf e}_3$ direction
with velocity $c$, the velocity of light in a vacuum. The bivector
$i S_{em}^{1/2}{\bf e}_3$ encaptures electromagnetic ${\cal E}$ and
${\cal H}$ fields perpendicular to ${\bf e}_3$. Since there is no
real component of the electromagnetic field vector, the field can
also not act directly on the real part of the electron's
wavefunction, or its density distribution. However, we found in the
analysis of the electron's acceleration in an external field that
the time derivative of the field components and the density
components are equal and of opposite sign. One may infer from that
finding that an external and oscillating field $S$ will lead to an
oscillating field $\rho$ via something akin to the Newtonian
principle: for every action (here, a variation of field intensity
due to an impinging electromagnetic field), there is a reaction
(here, a variation of mass density of the electron). To see whether
such an assumption is justified we consider an electromagnetic field
propagation in $z$-direction with the field vector $S_{em}$ given
by:
\begin{equation}
S_{em}(z,t) = S_0(t) \sin^2\left(\frac{2 \pi}{\lambda} z - 2 \pi \nu
t \right)
\end{equation}
Given that the field propagates with relativistic velocity we
assume, for this treatment, that the electron velocity is very small
and that in effect the propagating EM field interacts at a given
point with the electron, which perceives the presence of the field
more or less as a time-dependent variation. At a point $z_0$, $\phi_0 = (2 \pi/\lambda) z_0$ this means:
\begin{eqnarray}
\dot{S}_{em}(z_0,t) &=& \dot{S}_0(t) \sin^2 \left(\phi_0 - 2 \pi \nu
t \right)
\\
&-&
4 \pi \nu S_0(t) \sin \left(\phi_0 - 2 \pi \nu t \right)
\cos \left(\phi_0 - 2 \pi \nu t \right) \nonumber
\end{eqnarray}
If the effect of the electron's density is equal to the change of the external field
\begin{equation}
d \dot{\rho}_{el}(z_0,t) = - \dot{S}_{em}(z_0,t),
\end{equation}
then the density change at $z_0$, which we assume to be initially
constant and $\rho_0$, will be:
\begin{equation}
d \rho_{el}(z_0,t) = \rho_0 - S_0(z_0,t) \sin^2 \left(\phi_0 - 2 \pi
\nu t \right)
\end{equation}
According to the relation between density and field components this also leads to a change
of the internal potentials $S_{el}$ of the electron, described by:
\begin{equation}
d S_{el}(z_0,t) = S_0(z_0,t) \sin^2 \left(\phi_0 - 2 \pi \nu t
\right)
\end{equation}
The limit of energy adsorption from the electromagnetic field is reached, when the electron density
has reached the limit:
\begin{equation}
\rho_{el}(z_0, t_f) = \rho_0 - S_0(z_0,t_f) =  0
\end{equation}
While this seems a coincidence at first view -- the amplitude $S_0$ could in principle
be arbitrary and the electron adsorb less energy than required to reach this threshold -- on reflection
it is quite understandable. After all, the external field $S_{ph}$ will have been created by the reverse
effect, i. e. an electron being decelerated and emitting a photon in the process. Such an emission process
is equivalent to an adsorption process with a negative time coordinate. From a physical point of view
it is therefore symmetric.

However, it has to be acknowledged that this particular point deserves a more careful analysis, which has to
be given in future. The second point, which has not been treated is the transfer of frequency, due to the external
$S$ field, to a change in wavelength, or a local change of electron density and field amplitudes after adsorption
of energy. So far, this point has been omitted since we have analysed energy transfer processes at one particular
coordinate $z_0$ only. To determine the phase difference between different points at the moment, when the
electron acceleration is terminated ($t = t_f$), we calculate the density at the point $z + dz$. From the
equations derived so far it is clear that:
\begin{eqnarray}
\phi_0 - 2 \pi \nu t_f = \frac{\pi}{2} \qquad S_0(z_0, t_f) = \rho_0
\end{eqnarray}
Calculating now the density at $z_0 + dz$ and $t = t_f$ we get:
\begin{eqnarray}
\rho_{el}(z_0+dz,t_f) &=& S_0(z_0,t_f) \\
&-& S_0(z_0+dz,t_f) \sin^2\left(\frac{\pi}{2} + \frac{2\pi}{\lambda} dz\right) \nonumber \\
&=& S_0(z_0,t_f) - S_0(z_0+dz,t_f) \cos^2\left(\frac{2\pi}{\lambda} dz\right) \nonumber
\end{eqnarray}
Setting now:
\begin{equation}
S_0(z_0+dz,t_f) \approx S_0(z_0,t_f) = \rho_0
\end{equation}
we obtain for the amplitude of the density at the point $z_0+dz$ the result:
\begin{equation}
\rho_{el}(z_0+dz,t_f) = \rho_0 \sin^2\left(\frac{2\pi}{\lambda}
dz\right)
\end{equation}
The local distribution of the density is then equal to the local distribution
of a wave with velocity $v = h/m \lambda$. A similar result is obtained for the
field component of electron propagation. Since both components show the same properties
as an electron wave in motion with velocity $v$ it is safe to conclude that the adsorption
of the $S$ field leads to an acceleration of the electron until it reaches a velocity
which is corresponding to the frequency of the incident dynamic field.

As both components comply with the standard form for the wavefunction, we may write the
electron wavefunction after acceleration as:
\begin{equation}
\psi_S(v) = \rho^{1/2}(v) + i S^{1/2} (v) = \rho_0 \exp
i\left(\frac{2 \pi}{\lambda} z - 2 \pi \nu t\right)
\end{equation}
One may now analyze the problem of an electron in motion undergoing
accelerations due to impinging photons. Given that the wavefunction
is linear, a subsequent acceleration will change it in exactly the
same manner as the first one. This means, that one can actually
analyze electron-photon interactions from the viewpoint of the
conservation of energy and momenta. This is, how one commonly
accounts for Compton scattering \cite{compton23}.

\section{The hydrogen atom}\label{hydrogen}

It is straightforward to apply the Schr\"odinger equation of the
electron, developed in previous sections, to the problem of a
central potential and to restrict the problem to finding solutions
of the equation:
\begin{equation}
\left(- \frac{1}{2} \nabla^2 - \frac{1}{r} \right)\psi_S = i
\frac{\partial \psi_S}{\partial t} = \epsilon \psi_S
\end{equation}
However, even though we have shown how the frequency of the electron
changes in an external electrostatic field, we have done so only for
one particular value of the electrostatic potential. We have not
analysed, whether the electron can have a {\em single} energy value,
described by $\epsilon$, in an environment, where the electrostatic
potential varies within the mass distribution of a {\em single}
electron. In particular, since we expect in such an environment that
the frequency changes continuously. Within the present framework the
Schr\"odinger equation, which describes the energy levels within a
hydrogen atom, is due to a slightly different physical situation
than treated previously. This situation is captured in postulate 4:
In equilibrium, the energy density throughout the space occupied by
a single electron is invariant. It can be argued that such a
postulate is necessary to account for the -- experimentally observed
-- stability of hydrogen. Let us assume that the postulate is not
valid. We may then write the equation in the following way:
\begin{equation}
\left(- \frac{1}{2} \nabla^2 - \frac{1}{r} \right)\psi_S =
\epsilon(r) \psi_S
\end{equation}
If a particular infinitesimal density component $d\rho$ is found at
a radius $r$, its energy value is described by $\epsilon(r)$. Now if
$\epsilon(r)$ varies with the radius, for example becoming more
negative with decreasing $r$, then the system can lower its energy
if $d \rho$ migrates to a position closer to the nucleus. In this
case the general principle of energy minimization requires that the
electron density collapses to an infinitesimal shell around the
nucleus. Since this is not, what we observe, postulate 4 must be
correct. And in this case the energy value $\epsilon$ must be
constant throughout the system.

The standard solution for the non-relativistic hydrogen problem is the following
(we give its standard textbook format):
\begin{eqnarray}
\psi_{nlm}(r, \vartheta,\varphi) &=& C_{nlm} \exp\left(- \frac{2 r}{n a_0}\right)
L_{n-l-1}^{2l+1}\left(\frac{2r}{n a_0}\right) \nonumber \\
 &\times& P_l^m(\cos \vartheta) \exp(i m \varphi)
\end{eqnarray}
Here, $C_{nlm}$ is a normalization constant. Since the hydrogen atom
is neutral, the total electron charge within its shell must be equal
to unity. $a_0$ is the Bohr radius, or 5.292 $\times$ 10$^{-11}$m.
Writing the wavefunction in a more compact manner we may state:
\begin{eqnarray}
\psi_{nlm}(r, \vartheta,\varphi) &=& U_{nlm}(r,\vartheta)\, e^{i m
\varphi}
\end{eqnarray}
Density components, as analyzed in great detail in the preceding
sections, are reflected by the real components of electron
wavefunctions, while the imaginary components relate to the
electron's fields. This allows to decompose the wavefunction of the
hydrogen electron into:
\begin{eqnarray}
\rho^{1/2} &=& U_{nlm}(r,\vartheta) \cos m \varphi \nonumber \\
S^{1/2} &=& U_{nlm}(r,\vartheta) \sin m \varphi
\end{eqnarray}
Here, the important result is that in general the electron within a
hydrogen atom also possesses density components and field
components, related to its state of motion. Doran and Lasenby have
given a fully relativistic treatment of the hydrogen problem within
geometric algebra, based on previous work of Arthur Eddington
\cite{edd36,kil94}, and found that the general solutions have to be
described by multivectors. This indicates the same feature: a
presence of density components and field components (see p. 294 of
Ref. \cite{dor02}).

The field plane is perpendicular to the direction of ${\bf
e}_{\varphi}$. It should be kept in mind that this wavefunction
describes an electron without spin: the components $m \ne 0$ relate
to additional motion of the electron around its $z$-axis. Including
spin into this description requires to modify all states, also the
groundstate $m = 0$. As the groundstate solution of hydrogen is
radially symmetric, and as the spin-component of the wavefunction is
decoupled from the local component and has exactly two solutions, we
may account for spin by the component:
\begin{eqnarray}
\chi^{\pm} &=& \frac{1}{\sqrt{2}} \left(1
\pm {\bf e}_{\vartheta} \wedge {\bf e}_{\varphi}\right) \\
\psi_{nlm}^{\pm}(r, \vartheta,\varphi, \sigma) &=&
\frac{1}{\sqrt{2}} U_{nlm}(r,\vartheta)\, e^{i m \varphi} \left(1
\pm {\bf e}_{\vartheta} \wedge {\bf e}_{\varphi}\right)\nonumber
\end{eqnarray}
For the groundstate wavefunction $m = 0$ the result will thus be:
\begin{eqnarray}
\psi_{nl0}^{\pm}(r, \vartheta,\varphi, \sigma) &=&
\frac{1}{\sqrt{2}} U_{nl0}(r,\vartheta)\,\left(1 \pm {\bf
e}_{\vartheta} \wedge {\bf e}_{\varphi}\right)\nonumber
\end{eqnarray}
Writing the local part of the wavefunction as a multivector leads to
the following formulation:
\begin{equation}
\psi_{nlm} = U_{nlm}(r,\vartheta) \left(\cos m \varphi + {\bf e}_r
\wedge {\bf e}_{\vartheta} \sin m \varphi \right)
\end{equation}
The complete solution including spin then can be written as the
following multivector:
\begin{eqnarray}
\psi_{nlm}^{\pm} &=& \frac{U_{nlm}(r,\vartheta)}{\sqrt{2}}
\\ &\times& \left\{\cos m \varphi + i \left[ \left({\bf e}_{\varphi} \mp {\bf e}_{\vartheta}
\right)\sin m \varphi \mp {\bf e}_r \cos m \varphi \right]\right\}
\nonumber
\end{eqnarray}

\section{The hydrogen molecule and density functional
theory}\label{hydrogen-molecule}

A hydrogen molecule or a helium atom poses the additional problem to
reconcile postulate 4 with the existence of one coherent electron
density throughout the system. In principle, the problem can be
solved using standard DFT based on the Kohn-Sham equations
\cite{ks65}. However, the Kohn-Sham formulation of DFT seems, from
the outset, somewhat cumbersome, since it requires the
self-consistent calculation of every single--electron state in a
system. The computational effort required as the system size
increases then rapidly becomes prohibitive. This has led to
extensive research into the possibility of a so-called orbital-free
formulation of DFT \cite{wang05}. The key problem in this line of
research is the kinetic energy functional, which to date cannot be
described in a completely transferable manner. Given the preceding
presentation, it seems quite unlikely that a relation, based on
electron density alone, could actually account for all aspects of a
many-electron system. If magnetic properties play a role at all,
then according to this model they must enter the theoretical
description. This applies not only to magnetic material, but also to
non-magnetic solids, as the minimum energy state of the system is a
consequence of the existence of an equal number of spin-up and
spin-down electrons. However, the many--electron problem was encoded
by Levy, Perdew and Sahni (LPS) in 1984 in an equation based only on
the density \cite{lps84} (see Eq. \ref{lps-01}).

\subsection{Modified density equation including
spin}\label{mod-density}

Within the present context, there are two problems with the
equation: (i) It describes a bosonic system, since all properties,
related to the spin of electrons, are absent from the formulation.
(ii) It does not account for the difference in nature of field
components (a bivector) and mass components (a scalar). It is hard
to see, therefore, how this relation could be complete. However, it
should be possible, given the analysis contained in this paper, to
accurately describe a non-relativistic many-electron system in its
groundstate by the following coupled equations:
\begin{eqnarray}\label{of-bivec}
\Psi &=& \rho^{1/2} + i {\bf e}_S S^{1/2} \nonumber \\
\left[-\frac{1}{2} \nabla^2 + v_{ext} + v_{eff,0} + i {\bf e}_v v_{i}\right]\Psi &=& \mu \Psi \\
\rho_0 &=& \Psi^{\dagger} \Psi \nonumber
\end{eqnarray}
$v_{eff,0}$ combines the Hartree potential and the cohesive
potential defined in Section \ref{wave-1}, but does not contain
exchange-correlation potentials. Exchange and correlation potentials
encode the difference between non-interacting and interacting
electrons. In this equation electron interactions beyond $v_{eff,0}$
are part of the bivector potential $i {\bf e}_v v_{i}$, since they
are due to field-mediated interactions within the electron
distribution. The direction of the unit vector ${\bf e}_v$ and the
intensity of the potential $v_i$ will depend on the system under
consideration and have to be determined self-consistently. The
bivector part of the many-electron wavefunction, the product of unit
vector ${\bf e}_S$ and electromagnetic intensity $S^{1/2}$ are
equally system and location dependent and have to be determined by
self-consistent iterations. The equation couples field terms to
density terms also via a bivector potential. The resulting mass and
field components of the wavefunction $\Psi$ then give the
conventional mass density $\rho_0$. Even though both ${\bf e}_S$ and
${\bf e}_v$ will vary from one point to the next, the set of coupled
equations is still much simpler than the full many-body treatment of
a many electron system in standard theory.

The justification to write the wavefunction of a many-electron
system in this way rests on three results of previous sections: (i)
The wavefunction of any single --
 and extended -- electron complies with a Schr\"odinger equation. A
 system composed of an aggregate of more than one electron then
 should also comply with a Schr\"odinger--like equation. (ii) The
 construction of a wavefunction from scalar and bivector components
ensures that the $4\pi$--symmetry of the wavefunction will be
retained, which is the key property of the wavefunctions of
fermions. (iii) The interaction between field components and mass
components implies bivector potentials in a crystal field of
interactions. Without such a potential many-body effects could not
be included in the description.

One can decompose the equation into scalar and bivector components
to make transparent, in comparison with the LPS equation, what the
additional components actually are. For the scalar part we get:
\begin{equation}
\left[-\frac{1}{2}\nabla^2 + v_{ext} + v_{eff,0} -
\mu\right]\rho^{1/2} = {\bf e}_v \cdot {\bf e}_S \,v_i S^{1/2}
\end{equation}
The change is thus not simply a potential term of the same general
form as the effective or external potential. Such a behavior could
be encoded e.g. in a density dependent exchange correlation
potential. However, here we have an additional term which will act
as a source for the electron density distribution. This source term
depends on the direction of the bivector potential as well as the
field components of electron motion. It is thus clearly beyond
current models in DFT. The bivector equation is equally instructive.
It can be written:
\begin{eqnarray}
\left[-\frac{1}{2}\nabla^2 + v_{ext} + v_{eff,0}- \mu\right]i {\bf
e}_S &S^{1/2}& \\
= {\bf e}_v \wedge {\bf e}_S v_i S^{1/2} - i {\bf e}_v v_i
\rho^{1/2} \nonumber
\end{eqnarray}
This is, in essence, a vector-equation with purely imaginary
variables. It thus introduces directional effects into the density
equation, and couples density components and field components. Also
this equation is clearly beyond current formulations in DFT. In
imaginary components and replacing the wedge product by the cross
product (${\bf a} \wedge {\bf b} = i {\bf a} \times {\bf b})$, we
get:
\begin{eqnarray}
\left[-\frac{1}{2}\nabla^2 + v_{ext} + v_{eff,0}- \mu\right] {\bf
e}_S &S^{1/2}& \\
= {\bf e}_v \times {\bf e}_S v_i S^{1/2} - {\bf e}_v v_i \rho^{1/2}
\nonumber
\end{eqnarray}
It can also be seen that either the field component $S$ or the
bivector potential $v_i$ by itself will not alter the general
equation. Only the field components of electron energy and the
bivector potential of the solid {\em together} could give an
accurate account of the charge distribution for electrons with spin.
It is also clear, by comparing with the hydrogen problem, that the
potential $v_i$ is absent in the groundstate of the hydrogen
electron: it is thus genuinely of many-body origin.

In a slightly more compact form, which differentiates between vector
and scalar components of the equation, we may write the following
coupled set of equations, where the local dependency of components
has been made explicit for clarity:
\begin{eqnarray}
\Lambda ({\bf r}) &=& {\bf e}_S S^{1/2} \nonumber \\
\Pi({\bf r}) &=& {\bf e}_v v_i  \\
v_0({\bf r}) &=& v_{ext} + v_{eff,0} \nonumber
\end{eqnarray}
\begin{eqnarray} \label{hofer-eq}
&&\left[ -\frac{1}{2}\nabla^2 + v_{0}({\bf r}) - \mu\right]
\rho^{1/2}({\bf r}) =
\Pi({\bf r}) \cdot \Lambda({\bf r})  \\
&&\left[ -\frac{1}{2}\nabla^2 + v_{0}({\bf r}) - \mu\right]
\Lambda({\bf r}) + \rho^{1/2}({\bf r}) \Pi({\bf r}) = \Pi({\bf r})
\times \Lambda({\bf r}) \nonumber
\end{eqnarray}
In general, the formulation is still much simpler than present
formulations of many-body theory. The simplification comes from the
fact that the model wavefunction contains only four (the amplitudes
$\rho^{1/2}$ and $S^{1/2}$ and the vector of unit length ${\bf
e}_S$) independent variables, which can be mapped onto a grid in
real space. It is thus much easier to evaluate in a general
minimization problem even though it captures {\em all} physical
properties of the system. Using, for example, a real--space method
and a grid spacing of about 0.2 to 0.3 \AA, it should become
computational routine to simulate systems of more than one million
atoms: an increase of the current limit of DFT of a factor of about one
thousand. In this case, mesoscopic systems will come into the range
of {\em ab--initio} methods and biological systems should be
tractable by theoretical means.

The indispensable first step for such a theoretical framework will
be to determine the bivector potential and the effective potential
in a simple system, say a system of constant electron density. Only
after such a potential has actually been determined can the
framework be applied to more complex systems with variable electron
density. While it cannot be guaranteed that this approach will lead
to a reliable method for computing the physical properties of many
electron systems, it seems justified to develop its consequences and
to compare the results to standard methods. The gain in theoretical
efficiency, if successful, would be quite substantial.

\subsection{Many-body wavefunction}

The groundstate wavefunction of an $N$-electron system can be
written as:
\begin{equation}
\Psi_{MB} = \Psi\left({\bf x}_1,{\bf x}_2,{\bf x}_3, ... , {\bf
x}_N\right)
\end{equation}
Here, the variables ${\bf x}_i$ combine local (${\bf r}_i$) and spin
($\sigma_i$) variables. The wavefunction $\Psi_{MB}$ thus contains
at least $3N$ variables, the position vectors of all $N$ electrons,
${\bf r}_1$ to ${\bf r}_N$. The main condition imposed upon the
wavefunction due to the Pauli principle is that it must be
antisymmetric: an exchange of two variable (or two electrons) ${\bf
x}_i$ and ${\bf x}_j$ ($i$ and $j$) changes the sign of the
wavefunction from plus to minus and vice versa. A general form of
the wavefunction complying with this condition is its construction
via Slater determinants \cite{slater29}:
\begin{equation}
\Psi_{MB} = \sum_{\kappa}C_{\kappa}
Det|\psi_1(x_1),\psi_2(x_2),\psi_3(x_3), ... ,\psi_N(x_N)|
\end{equation}
For a hydrogen molecule the many-body wavefunction of the system,
including the spin functions $\alpha$ and $\beta$, is given by
\cite{lev04}:
\begin{equation}
\Psi({\bf r}_1,{\bf r}_2,\sigma_1,\sigma_2) = C
\left|\begin{array}{cc}
  \psi_{100}({\bf r}_1)\alpha(\sigma_1) & \psi_{100}({\bf r}_2)\alpha(\sigma_2) \\
  \psi_{100}({\bf r}_1)\beta(\sigma_1) & \psi_{100}({\bf r}_2)\beta(\sigma_2)
\end{array}\right|
\end{equation}
The total (groundstate) wavefunction is therefore:
\begin{eqnarray}
\Psi({\bf r}_1,{\bf r}_2,\sigma_1,\sigma_2) &=& C \psi_{100}({\bf
r}_1) \psi_{100}({\bf r}_2)\nonumber \\ &\times& \left[
\alpha(\sigma_1)\beta(\sigma_2) -
\beta(\sigma_1)\alpha(\sigma_2)\right]
\end{eqnarray}
Since a hydrogen molecule in its groundstate does not interact with
a magnetic field, one may omit the spin-components of the
wavefunction and write:
\begin{equation}
\Psi_{H_2}({\bf r}) = \psi_{100}({\bf r} - {\bf R}_1)
\psi_{100}({\bf r} - {\bf R}_2) = \rho_0^{1/2}({\bf r})
\end{equation}
where we have assumed that the nucleus of the first atom is at ${\bf
R}_1$ and the nucleus of the second atom at ${\bf R}_2$. However,
including spin becomes necessary if one considers the gradual
transition from the hydrogen molecule to two hydrogen atoms
\cite{gunnarson74}. In this wavefunction we may account for spin in
a similar way as for the hydrogen atom, with the only difference
that now we have two centers which serve as the origin of radial
unit vectors. These unit vectors are symbolized by ${\bf e}_r^1$ and
${\bf e}_r^2$, respectively.
\begin{eqnarray}
\Psi_{H_2}({\bf r},\sigma) &=& \frac{1}{2} \psi_{100}({\bf r} - {\bf
R}_1) \psi_{100}({\bf r} - {\bf R}_2) \\
&\times& \left[\left(1 + i {\bf e}_r^1\right)\left(1 - i {\bf
e}_r^2\right) - \left(1 - i {\bf e}_r^1\right)\left(1 + i {\bf
e}_r^2\right)\right]\nonumber
\end{eqnarray}
In its triplet state the spin component of the two electron
wavefunction is symmetric, while the local component becomes
antisymmetric. In this case one has to construct the charge and spin
densities from the $1s$ and $2s$ orbitals, respectively.

\section{Discussion}\label{discussion}

We have on purpose avoided {\em any} reference to classical
mechanics in this work on electrons. The reason is that we think
that mechanics has obscured the meaning of certain formulations in
quantum mechanics. In that respect, we also think that the name {\em
quantum mechanics} carries a wrong message, and that it is better to
talk about either wavedynamics or microdynamics, although this might
be a secondary issue. The primary issue, we hope to have clarified
in this paper, is that electrons are not mechanical objects. This,
of course, is well known. However, so far no detailed account
existed, how density and spin properties of single electrons might
be distributed in space and time {\em without any measurement}. We
think that the concept of an extended electron, introduced here,
might be easier to reconcile with experimental results at the atomic
scale, with density functional theory, and with visualizations of
physical processes. It is often the picture in the mind, which leads
to new insights, and not the mathematical formalism. Whether this
electron model describes the {\em real} electron or not, is a moot
point. As demonstrated, it leads to the same numerical results as
the standard model, it accounts for most experimental results
obtained so far, it gives a consistent interpretation of the
electron wavefunction, and it allows to picture atomic-scale
processes in space and time. Even if it were not {\em real}, this
electron seems quite a {\em useful} model.

It is instructive to look at the presented material from the
viewpoint of basic principles. Quite apart from the historical
development of modern electron theory, this should allow us to
appreciate the importance or unimportance of specific findings. The
first fundamental finding, and which made a wave theory of electrons
initially necessary, is actually electron diffraction. That this was
initially a hypothesis in a PhD thesis of Louis de Broglie, makes it
all the more remarkable. In a modern context, one could say it is
the observation of standing waves on a metal surface, which leaves
no other conclusion but to assign wave properties to electrons. In
this respect it is hard to overestimate the importance of the
experiments coming out of Don Eigler's lab in the early 1990s.

As a second fundamental finding one has to nominate photoelectric
currents as interpreted by Einstein with the help of Planck's
hypothesis. Here, it is found that the frequency of the electron
waves can be altered by electrostatic fields, as existing within a
metal crystal. This finding paves the way for the formulation of the
initial Schr\"odinger equation, which can be used to determine the
emission frequencies of hydrogen atoms, admittedly one of the great
problems of 19th century physics. However, this finding alone does
not allow for a treatment of interactions with electromagnetic
fields.

Here, we need the third fundamental finding, which we consider to be
the Aharonov-Bohm effect. Only in this context can it be understood
that the vector potential changes the wavelength and thus has to be
treated differently than, e.g. the changes in an electrostatic
potential, where the frequency is affected. It is also quite
interesting that this effect was a theoretical prediction based on
the Schr\"odinger equation and formulated long after the equation
had been used to solve the hydrogen problem. The Aharonov-Bohm
effect is as important as the photoelectric effect. Without it, the
only way to include vector potentials in the Schr\"odinger equation
would be via a classical Hamiltonian $H = \left[{\bf p} - (e/c) {\bf
A}\right]^2$ \cite{jackson-ed}, and the correspondence principle.
This, in turn, would require to treat the electron as a classical
point-particle and lead to inconsistencies in the suggested
theoretical model.

The most intriguing part of the framework seems to concern the
possibility of constructing wavefunctions from density and field
components and, by the reverse process, of decomposing wavefunctions
into density and field components. This should add substantial
physical insight to any theoretical result obtained e.g. within a
many-body framework. In the final analysis we think that the role
and the actual  -- i.e. physical -- content of the wavefunctions
have not been sufficiently clear. An example of this lack of clarity
in our view is the famous Schr\"odinger--cat thought experiment
\cite{schr35}. There, a superposition of two states, a living cat
and a dead cat, collapses into one final state upon measurement,
i.e. the interaction of an electron with an external field. The
important feature of the experiment lies in its statistical
interpretation of the wavefunction. As we do not know, which state
the electron is in, the cat could be said to be dead and alive at
the same time -- which seems quite impossible for a living organism.
However, as shown in the treatment on spin and its interaction with
a magnetic field, the electron wavefunction within the present
framework will always indicate a well defined physical state, as
described by time and space dependent density and field components.
The statistics in this case do not enter via some unknown "quantum"
property of electrons, but are due either to a variation of the
electron's properties in space and time (which, for example,
prevents us seing the density variations of a surface--state
electron if it is not scattered on a defect, and makes us detect
standing waves at a step edge or a single atom \cite{eigler93}), or
to a lack of knowledge about the individual electron (which makes us
detect two deflection spots in a Stern-Gerlach experiment
\cite{stern22}). In this case, if the life of a cat is related to
one particular electron state, then this cat is definitely dead or
alive, but never both. If we have to describe it as a superposition
of alive and dead, then we have to concede that we do not know
exactly which electron we are talking about, or at which precise
moment our measurement is taken. This is, in essence, the same lack
of information encountered in statistical thermodynamics.
Incidentally, this view is very close to the one advocated by Erwin
Schr\"odinger \cite{schr35}.

However, we have to concede that the model does not provide a
comprehensive theory for the observed self-interference of single
electrons \cite{donati73}. Interference here is due to the spacetime
variations of the electron's wavefunction and its geometric
characteristics upon scattering. How a single electron could change
its wave pattern by interacting with an environment which is well
separated from any one of its possible trajectories, cannot be
answered within this model. The only way, it seems, that such a
behavior could be reconciled with the present model, is by invoking
a detection loophole. If, for example, the impacts detected depend
on a threshold energy of the impinging density wave {\em and} the
state of the atoms making up the detector, then it seems possible
that detection is actually a very rare and, due to the excitations
of the detector atoms, essentially {\em stochastic} event and that
most density waves remain undetected, but do contribute to the
interference pattern. That such a scenario is not as inconceivable
as it might seem, can be inferred from scanning tunneling microscopy
experiments, where only a tiny fraction, or 10$^{-5}$ of the actual
current in the tunnel junction is detected in the experiments
\cite{hofer03}.

In a wider context, there is a long tradition in experimental
physics, starting with Alain Aspect's experiments in the 1980s
\cite{aspect81}, on the non-locality of quantum mechanics,
formalized in the Bell inequalities \cite{bell64,bell87}. These
experiments remain beyond the scope of this presentation, although
it cannot at present be excluded that non-local correlations could
play a role in the properties of the bivector potential in
many-electron systems. Given the role of non-local correlations, for
example in the construction of kinetic energy density functionals
\cite{perdew09}, such a possibility should certainly not be
excluded. From a different point of view, it can be said that the
Bell inequalities \cite{bell64} are violated, whenever a system
violates at least one of following two conditions: (i) Its
interactions are strictly local, and(ii) its spin is counterfactual
definite (see Ref. \cite{blaylock09} and references therein). {\em
Counterfactual definite} means that an electron possesses a
particular \mbox{(= definite)} spin, regardless of whether it is
measured or not. Quantum mechanics, for example, violates both of
these conditions: the first, in the collapse of the wavefunction
(which is non--local), the second , in the dependence of spin on a
measurement (as spin is generally isotropic without a measurement).
This question requires a very detailed and careful analysis, which 
shall be pursued in a follow--up paper.

From the viewpoint of condensed matter theory and computational
methods the extension of the equation derived by Levy, Perdew, and
Sahni \cite{lps84}, via a distribution of density and field
properties, given in Eq. (\ref{hofer-eq}), is certainly the most
important aspect of the model. It seems at first view surprising
that one may describe systems composed of  a large number of
electrons by a wavefunction with only four variables, i. e., the
amplitudes of density and field components, and the direction of the
field. In particular, since the standard model in many--body physics
requires at least $3N$ variables. While the final proof that the
model is suitable also for many-body systems will have to come from
the application of the model, its predictions, and its agreement
with experimental measurements, it can be analyzed, where the actual
gain in efficiency could potentially come from. From the viewpoint
of DFT, the key problem in orbital free models always was the
kinetic energy functional \cite{wang05}. In the Kohn-Sham
formulation of DFT, this problem is transferred to the
exchange-correlation functionals, which describe the difference in
kinetic energy between interacting and non-interacting electrons
\cite{ks65}. In the present model, the kinetic energy density is
given by the Laplace operator acting on the density and field
components of electrons, e. g. in a solid. It is thus different from
the von Weizs\"acker or the Thomas-Fermi kinetic energy densities
\cite{wang99}. However, as shown in the derivation of the free
electron and the hydrogen atom, it reverts to these limiting cases
with a vanishing bivector potential in case of single electrons.

In many-body theory, the many-body aspect is usually encoded in the
construction from single-electron states; here, the many-body aspect
is accounted for by the bivector potential and by the phase
coherence of densities and field components. The present model of a
many-body system in a sense does not know of single, i.e., isolated
electrons, since every single electron is connected to adjacent
electrons via coherent density functions and fields. In the present
model, the many-body aspect of a system is thus encoded in its
coherence, and in its common energy or chemical potential $\mu$.
Since it does not know of single electron states, it also does not
require to build up the many-body wavefunction from Slater
determinants defined via a $3N$-variable space. It is thus much
closer to DFT than existing many-body formulations, even though
electron correlations within solids should be accurately described.
Considering, that phase coherence of $\Psi$ also plays a role across
the boundary of a periodic system, most principles known to
condensed matter theorists which come from periodic boundary
conditions should be applicable. However, since the potential
efficiency is much higher, in particular in the description of large
systems, it should also be possible, without lack of precision, to
simulate very large systems, i.e. mesoscopic systems and mesoscopic
timescales.

DFT continues to be a project in development. In essence, its
successful application to most problems in solid state physics and
physical chemistry has made it an indispensable tool of theorists
today. However, the current efforts to determine the kinetic energy
of a many electron system \cite{perdew09,burke10} and to go beyond
Kohn-Sham theory also reveal that the present theoretical framework
is not an endpoint in the development. One of the motivations to
develop this theoretical model of electrons was the attempt to
understand, from the very principles of modern physics, why an
electron appears such a strange entity. The hope was that a better
understanding of fundamental processes and effects may lead to a
better formulation of the many-electron problem. The bivector
potential $\Pi$, introduced in this paper, is a genuinely novel
concept. It arises naturally, as seen, if one tries to generalize
the LPS equation for a spin system. What the exact form of this
potential is and how it can be formalized in terms of
spin-densities, remains to be seen.

We have briefly mentioned in this paper that a relativistic
theoretical framework, based on the Dirac equation and geometric
algebra, has been developed by Hestenes and others
\cite{hest84,dor02}. However, we have not extended the present
framework to the relativistic domain. A reason for this omission is
that a non-relativistic framework seems to be sufficient for the
time being, in particular in view of applications in condensed
matter. In this respect it is also worth reflecting the fact that
the Schr\"odinger equation, contrary to the wave equation, is
actually Galilei-invariant (under a kinematic transformation of the
wave function, see Ref. \cite{jackson-ed} p. 516). Which means that
it remains unchanged for systems at low velocity, which applies to
practically all systems studied under laboratory conditions. Under
these conditions the first postulate of the theory of relativity
remains thus valid, while the second postulate is also valid, since
the speed of light is much larger than the velocity of electrons.
Imposing an additional Lorentz-invariance on the systems under
consideration in DFT seems thus unnecessary at this point. The only
real motivation for developing such a Lorentz-invariant model of
electrons for applications in condensed matter physics would be, if
the description in this manner could be made substantially simpler.
At present, this seems not the case. However, we shall return to
this problem in future publications.

\section*{Acknowledgements}
This paper captures the essence of about fifteen years of research
into electron properties from three perspectives: standard quantum
mechanics, density functional theory, and experiments at the atomic
scale. It contains thus all the increments in understanding brought
about by discussions with colleagues over the years. Of all the
colleagues I should thus acknowledge I can name only one: Jaime
Keller, who introduced me to geometric algebra. The draft of this
paper was carefully read by Iain R. McNab and Krisztian Palotas, who
I have to thank for their contribution and their valuable
suggestions. I also have to thank Air France: it was on a night
flight from Paris to Beijing, when the first ideas to this concept
crystallized in my mind. And finally, I have to thank the Royal
Society for continued financial support.

\end{document}